\documentclass[prx,two column,showpacs,preprintnumbers,amsmath,amssymb,revtex4-1]{revtex4-1}
\usepackage{amsfonts, amsmath, bm, bbm, graphicx, epsfig, epstopdf}
\usepackage{dcolumn}
\usepackage{textcomp}
 \usepackage[caption=false]{subfig}
\usepackage{mathtools}
\usepackage{multirow}   
\usepackage{booktabs}  
\usepackage{graphicx}

\usepackage{xcolor}
    \usepackage{braket}
\renewcommand\bra[1]{{\langle{#1}|}}
\makeatletter
\renewcommand\ket[1]{%
  \@ifnextchar\bra{\k@t{#1}\!}{\k@t{#1}}%
}
\newcommand\k@t[1]{{|{#1}\rangle}}
\makeatother

\begin{document}

 \title{Blockade-induced exchange primitives for scalable neutral-atom QPU}

\author{Mohammadsadegh Khazali$^1$}
\email{mskhazali@ut.ac.ir}
\author{Klaus M\o lmer$^2$}

\affiliation{%
$^1$Department of Physics, University of Tehran, Tehran, Iran \\
$^2$Niels Bohr Institute, Jagtvej 155 A, 2200 Copenhagen, Denmark
}

\maketitle

\section*{abstract}

{\bf Many quantum hardware platforms natively support either phase or exchange operations, yet converting between these two forms of control typically incurs substantial overhead. Rydberg-blockade neutral-atom arrays are highly developed for phase control. Here, controlled exchange is introduced as a native, blockade-programmed phenomenon in a collective excited manifold. Target operations are engineered such that two competing exchange pathways between $\ket{01}$ and $\ket{10}$ destructively interfere, while a single collective four-photon channel mediated by a symmetric Rydberg excitation remains resonant and drives a direct SWAP, with all other qubit configurations undergoing an identity action. Exchange conditionality follows from blockade, where exciting a control atom to a Rydberg state shifts and blocks the target collective resonance, suppressing exchange, whereas leaving the control in the ground manifold enables exchange in a single step. This approach yields a family of controlled-exchange gates with coherent average gate fidelities above 99.99\% at selected operating points, while reducing circuit depth and Rydberg-state exposure time by an order of magnitude relative to decomposed implementations. Joint parameter optimization reduces sensitivity to laser-intensity fluctuations, Doppler shifts, and interatomic-distance uncertainty, with several SWAP-family gates retaining fidelities above 99.6\% under realistic experimental fluctuations and Rydberg decay. Finally, the anisotropic interaction generalizes the same principle to multiplexed controlled-exchange operations, providing compact hardware-level primitives for conditional information routing in extended neutral-atom arrays. }

\section*{Introduction}

The practical power of a quantum processor is shaped not only by coherence and scale, but also by which operations it can perform natively. 
When key primitives must be compiled from an ill-matched native set, circuit depth and error accumulation can quickly dominate performance.
Rydberg-blockade neutral-atom arrays now offer large, programmable registers with strong, controllable interactions and have enabled both many-body simulation and scalable quantum processing \cite{Saf10,Ebadi21,Semeghini21,Blu24}. 
Their native entangling toolbox is exceptionally well developed for controlled-phase operations, from foundational blockade-gate proposals to high-fidelity dark-state and parallel multiqubit implementations \cite{Jaksch00,Moller08,Pet17,Levine2019,Kha20,Eve23}. 

A complementary primitive, however, remains costly on this platform, namely \emph{controlled exchange}. SWAP-family operations are a workhorse for state comparison (overlap and purity estimation) \cite{Buhrman1999,Eke02}, for realizing exchange Hamiltonians in simulation \cite{Bloch2012}, and for routing information in architectures with geometric constraints \cite{Linke2017}. 
On neutral-atom processors, these operations are commonly decomposed into phase gates and single-qubit rotations, inflating depth and increasing exposure to Rydberg-state loss mechanisms \cite{Shende2004}.

 \begin{figure}
\centering
\includegraphics[width=\linewidth]{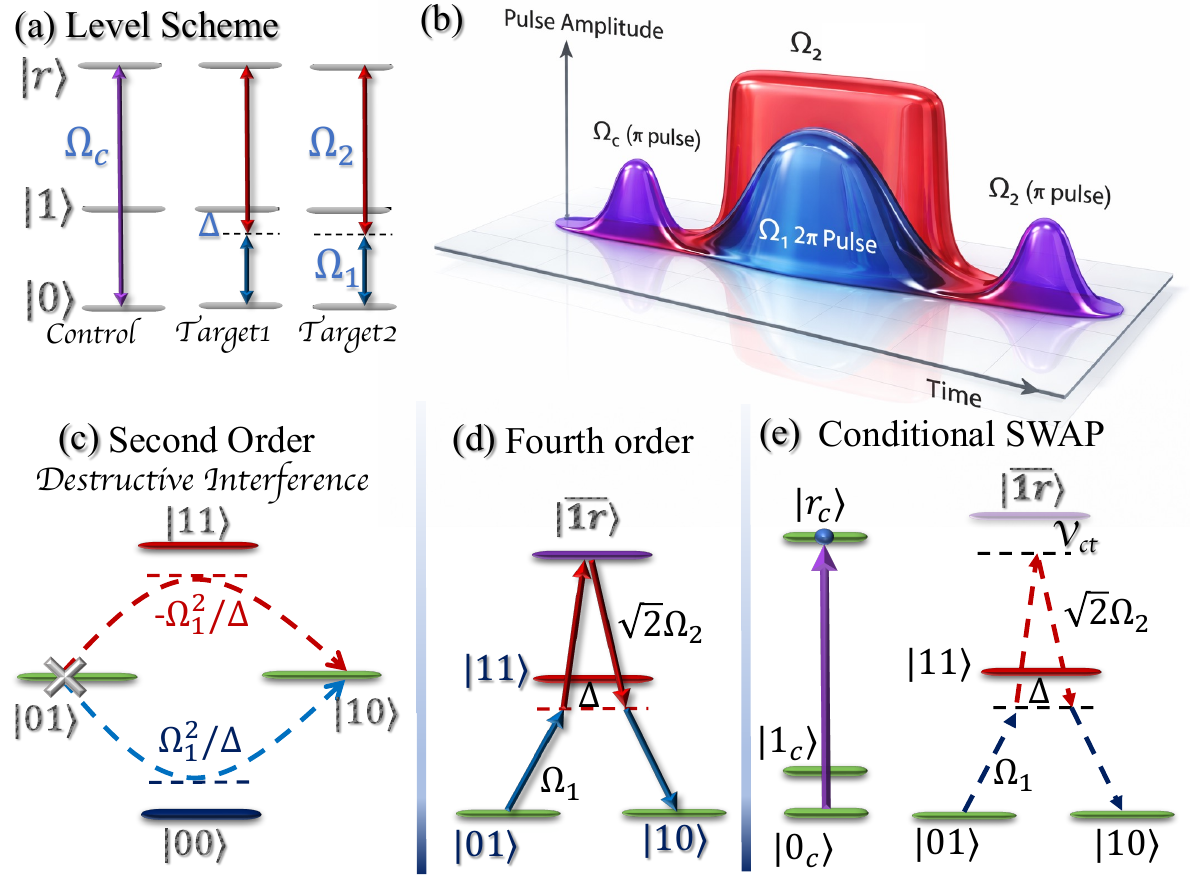}
\caption{{\bf Blockade-programmed mechanism for SWAP and controlled-SWAP.
(a)} Control–target level structure. {\bf (b)}  Pulse sequence: the control atom is transiently excited/deexcited to the Rydberg state at the beginning/end to make the intermediate SWAP operation, performed by $\Omega_{1,2}$, conditional.
 {\bf (c)} Direct (second-order) exchange $|10\rangle \rightleftarrows |01\rangle$ cancels because the two pathways via $|00\rangle$ and $|11\rangle$ carry opposite-sign detunings and interfere destructively. {\bf (d)} Driving the Rydberg transition with $\Omega_{2}$ opens a resonant four-photon exchange channel via the symmetric collective excitation $|\overline{1r}\rangle=(|1r\rangle+|r1\rangle)/\sqrt{2}$, enabling SWAP, while other computational configurations remain dark under $\Omega_{1,2}$ (Fig. 2). {\bf (e)} With the control in $|r\rangle$, control–target blockade shifts and blocks the collective channel, turning exchange off and producing a controlled-SWAP; anisotropic interactions can selectively blockade chosen target pairs for multiplexed controlled exchange (Fig. 5).}\label{Fig1}
\end{figure}

Here we introduce a blockade-enabled exchange mechanism in which coherent SWAP operations arise not from direct coupling, but from the {\it selective survival of a collective pathway protected by interference and blockade}. 
At the microscopic level, an off-resonant laser transition between the logical states $|0\rangle$ and $|1\rangle$ generates two nominal exchange pathways between $|01\rangle$ and $|10\rangle$, mediated by the intermediate states $|00\rangle$ and $|11\rangle$ with opposite detunings. These contributions cancel exactly due to destructive interference caused by the opposite phase generated by opposite detunings. By introducing a dipole coupling from $|1\rangle$ to a Rydberg state $|r\rangle$, a higher-order exchange channel opens via the symmetric collective state $(|1r\rangle + |r1\rangle)/\sqrt{2}$ generated by the intra-target Blockade effect. All other qubit configurations evolve under dark state or off-resonant phase-adjusted transitions, experiencing an identity operation.

The combination of long-range Rydberg interaction and the Rydberg collective pathway nature of this exchange provides a natural route to conditional control. When additional control atom(s) are excited to a Rydberg state, the control-target Rydberg blockade suppresses the collective exchange channel of the target atoms, preventing the SWAP operation. The exchange, therefore, occurs only when the control atom(s) remain in their ground manifold, realizing a controlled-SWAP gate. Exploiting the anisotropy of Rydberg interactions further enables state-dependent selection of exchange partners, offering a native mechanism for coherent conditional routing and quantum random-access memory within a single neutral-atom array.

Very recent studies have proposed high-fidelity iSWAP and exchange gates based on resonant dipolar interactions between distinct Rydberg states, with pulse shapes determined by optimal control \cite{Kha26,Ber25,Ild26}. Here, an explicit effective model of blockade-induced exchange resolves the interfering virtual pathways and their light shifts, allowing us to analytically formulate the physics and quantify sensitivity to Doppler shifts, laser-intensity fluctuations, and interaction uncertainties. These analytical relations guide the joint optimization of coupling ratios, detunings, interactions, and pulse parameters, identifying robust operating regimes while limiting errors from Rydberg spontaneous emission. Selected SWAP and controlled-SWAP variants achieve average gate fidelities above $99.6\%$ under realistic experimental fluctuations. Earlier proposals for direct Rydberg controlled exchange have exploited antiblockade to mediate transitions through the doubly excited state $|rr\rangle$ \cite{Wu21, Wan25}. In resonant implementations, distance-dependent variations of $V(d)$ perturb the pair-excitation resonance. Detuning errors comparable to the effective two-photon coupling can degrade fidelity even when small relative to the single-atom Rabi frequencies. Although detuned antiblockade variants improve robustness, our protocol generates exchange through blockade-induced interference in the regime $|V|\gg\Omega$, keeping double excitation energetically suppressed and operating away from pair-state resonances. Native implementation also reduces circuit depth and time-integrated Rydberg population by approximately an order of magnitude relative to decomposed constructions.



\section*{Results}

\subsection*{Controlled-SWAP scheme} 
 The level scheme of C-SWAP, shown in Fig.~\ref{Fig1}, consists of logical states $|0\rangle$ and $|1\rangle$ corresponding to hyperfine ground states, while $\ket{r}$ denotes a high-lying Rydberg level. 
 The hyperfine states $|0\rangle$ and $|1\rangle$ are coupled by a global microwave field  \cite{Che25,Wei11} or a local Raman transition \cite{explanation2} with Rabi frequency $\Omega_1$  and detuning $\Delta$.  The Rydberg-exciting optical fields $\Omega_c$ and $\Omega_2$ address the control and target atoms, respectively. Target selection is thus provided by the Rydberg-excitation field $\Omega_2$, without requiring an additional beam to shift the hyperfine transition frequencies; see below. The gate proceeds through the following three steps:\\
(i) {\it Control excitation:} A $\pi$ pulse  transfers the control atom from $\ket{0}$ into the Rydberg state $\ket{r}$.\\
(ii) {\it Target driving:} Simultaneously, the two target atoms are driven by fields $\Omega_1$ and $\Omega_2$, coupling $\ket{0}\rightarrow \ket{r}$ via (but detuned from) the other qubit state $\ket{1}$.  When the control atom occupies $\ket{r_c}$, the Rydberg–Rydberg interaction shifts the target levels, engineering destructive interference that blocks the SWAP. Conversely, if the control remains in $\ket{1}$, the target system undergoes the SWAP operation.\\
(iii) {\it Control retrieval:} A second $\pi$ pulse returns the control atom from $\ket{r}$ back to $\ket{0}$, completing the gate.\\

   \begin{figure*}
\centering
\includegraphics[width=\linewidth]{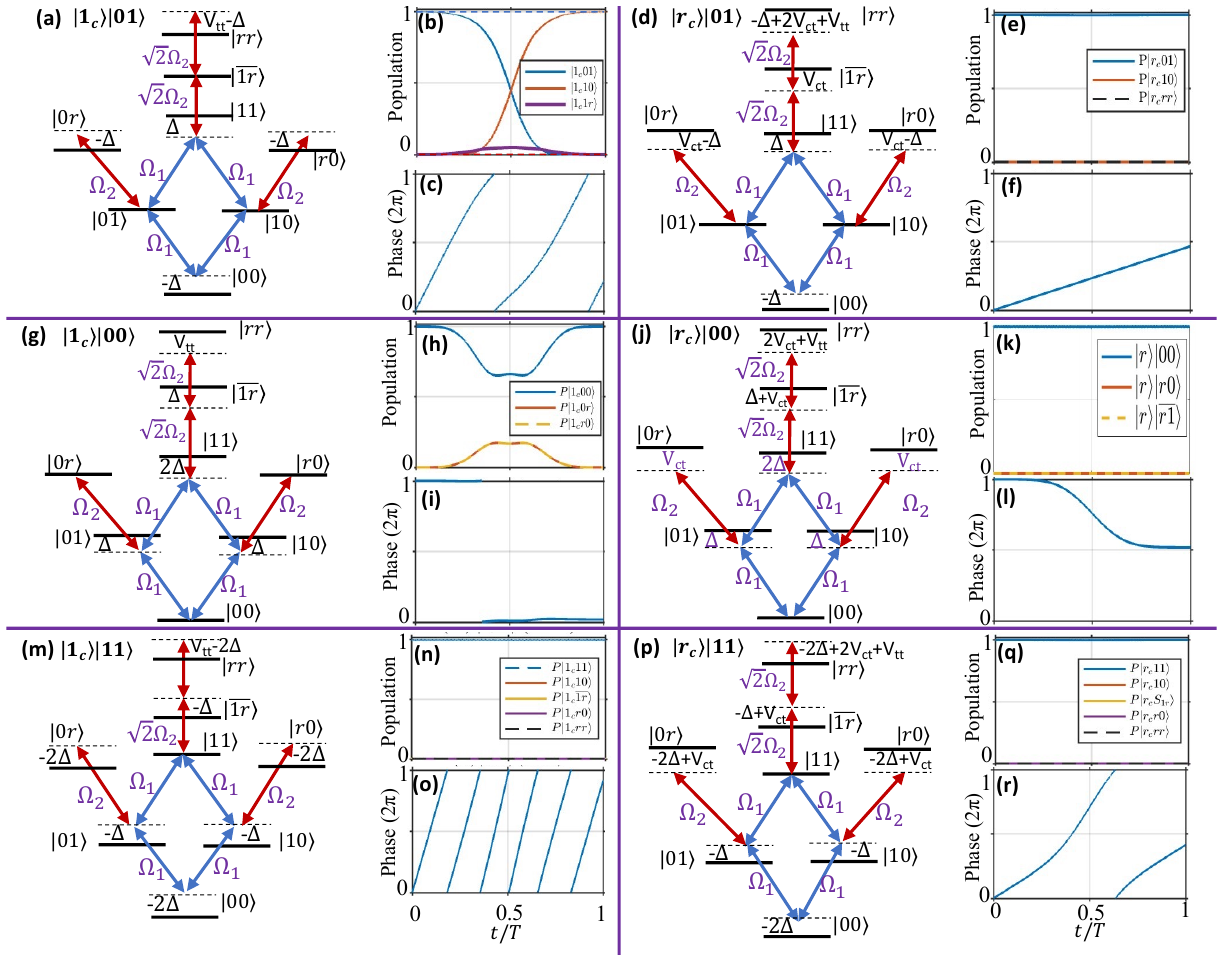}
\caption{{\bf Controlled-iSWAP Operation} -- Each group of three panels shows the collective target level scheme, target populations, and accumulated phase during target driving in step~(ii), in that order. The six input configurations are shown in (a)--(c), (d)--(f), (g)--(i), (j)--(l), (m)--(o), and (p)--(r). (a), (b) In the absence of control Rydberg excitation, the four-photon transition via the symmetric state $|\overline{1r}\rangle$ enables the SWAP between $|01\rangle$ and $|10\rangle$.  Note that it is essential that neither of the target atoms is Rydberg-blocked by the control atom. The dashed lines in (b) correspond to a case where one of the targets is blocked by the control under a non-isotropic interaction, illustrating that the SWAP operation cannot proceed via $|1r\rangle$ or $|r1\rangle$ alone but requires the symmetric state $|\overline{1r}\rangle = (|1r\rangle + |r1\rangle)/\sqrt{2}$.
The same principle will be employed for a conditional multiplexed SWAP operation at the end of the paper. (d), (e) When the control atom is in the Rydberg state $|r\rangle$ and targets are in $|10\rangle$ or $|01\rangle$, the interaction with the target atoms shifts the energy levels such that two eigenstates of Eq.~\eqref{Eq_H} become nearly degenerate. This degeneracy leads to {\it destructive interference}, effectively blocking the SWAP between target atoms. 
 (g) For the input $|1_c00\rangle$, the resonant two-photon
pathways to $|0r\rangle$ and $|r0\rangle$ admit the dark-state
description discussed in the text. At the pulse parameters
used here, bright-state admixture modifies the transient
populations and can contribute to the phase. Panels (h)
and (i) show the population and phase evolution obtained
from the full Hamiltonian. For the configurations in
panels (j), (m), and (p), off-resonant excitation largely preserves
the initial populations while dynamical phases accumulate,
as shown by the population traces in (k), (n), and (q) and the phase traces in (l), (o), and (r).
Applied parameters are presented in the 9th row of Table \ref{tab_1}.
The level diagrams highlight the principal exchange and light-shift pathways. Virtual excitation of $|rr\rangle$ produces a finite shift of the symmetric $|\bar{1r}\rangle$ state, which would be compensated by parameter adjustments in Table.~I.}\label{Fig2}
\end{figure*}

During step~(ii), the two target atoms are governed by the
following rotating-frame Hamiltonian, written for real Rabi
frequencies with $n_{\alpha i}=\ket{\alpha_i}\bra{\alpha_i}$
denoting the projector onto state $\ket{\alpha_i}$ of target atom $i$:
\begin{equation}\label{Eq_H}
\begin{aligned}
H_t={}&\sum_{i=1}^{2}\left\{
\frac{\Omega_1}{2}|1_i\rangle\langle0_i|
+\frac{\Omega_2}{2}|r_i\rangle\langle1_i|+\mathrm{h.c.}\right.\\[-1mm]
&\left.\hspace{19mm}+\Delta n_{1i}+\delta_r n_{ri}\right\}
+V_{tt}n_{r1}n_{r2}.
\end{aligned}
\end{equation}
Here $\delta_r=V_{ct}$ for a Rydberg-excited control and $\delta_r=0$
for a control in $|1_c\rangle$. The symmetric state
$|\overline{1r}\rangle=(|1r\rangle+|r1\rangle)/\sqrt2$
couples to both $|11\rangle$ and $|rr\rangle$ with matrix element
$\Omega_2/\sqrt2$. Although blockade suppresses double excitation,
virtual occupation of $|rr\rangle$ shifts the $|\overline{1r}\rangle$ collective state and
therefore affects exchange.

\begin{table*}[t]
\centering
\scriptsize
\setlength{\tabcolsep}{1.5pt}
\renewcommand{\arraystretch}{1.15}
\newcommand{\NoisePair}[2]{\shortstack{#1\\#2}}
\resizebox{\textwidth}{!}{%
\begin{tabular}{@{}rlc*{8}{c}cc*{5}{c}c@{}}
\hline
\multicolumn{3}{c}{Operating point}
& \multicolumn{8}{c}{Optimized gate parameters}
& \multicolumn{2}{c}{Nominal performance}
& \multicolumn{5}{c}{Excess infidelity by noise source (\%)}
& \multicolumn{1}{c}{Joint} \\
\cline{1-3}\cline{4-11}\cline{12-13}\cline{14-18}\cline{19-19}
\# & Gate & \shortstack{Temp.\\($\mu$K)}
& \shortstack{$T$\\($\mu$s)}
& $\sigma/T$
& \shortstack{$\Omega_{1,\max}/2\pi$\\(MHz)}
& \shortstack{$\Omega_2/2\pi$\\(MHz)}
& \shortstack{$\Delta/2\pi$\\(GHz)}
& \shortstack{$V_{tt}/2\pi$\\(GHz)}
& \shortstack{$V_{ct}/2\pi$\\(GHz)}
& \shortstack{Phase\\correction}
& \shortstack{$F_{\mathrm{coh}}$\\(\%)}
& \shortstack{Rydberg\\loss (\%)}
& $\delta I_1$ & $\delta I_2$ & Doppler
& $\delta V_{tt}$ & $\delta V_{ct}$
& \shortstack{$F_{\mathrm{tot}}$\\(\%)} \\
\hline
1 & SWAP & 2
& 1.27 & 0.23 & 63.4 & 70.6 & 1.02 & 0.69 & --
& $1.8^\circ[t_{1,2}]$ & 99.86 & 0.02
& \NoisePair{0.0006}{0.003} & \NoisePair{0.0004}{0.017}
& \NoisePair{0.005}{0.008} & \NoisePair{0.14}{0.04}
& -- & 99.63 \\
2 & SWAP & 150
& 0.697 & 0.31 & 85.8 & 234.78 & 1.51 & 1.017 & --
& $1.8^\circ[t_{1,2}]$ & 99.79 & 0.005
& \NoisePair{0.01}{0.03} & \NoisePair{0.01}{0.25}
& \NoisePair{0.08}{0.005} & \NoisePair{0.04}{0.002}
& -- & 99.36 \\
\hline
3 & iSWAP & 2
& 1.75 & 0.19 & 49.18 & 94.56 & 1.44 & 0.814 & --
& $18.1^\circ[t_{1,2}]$ & 99.997 & 0.016
& \NoisePair{0.007}{0.025} & \NoisePair{0.0003}{0.047}
& \NoisePair{0.0017}{0.002} & \NoisePair{0.1}{0.03}
& -- & 99.76 \\
4 & iSWAP & 150
& 1.74 & 0.19 & 49.23 & 96.11 & 1.45 & 0.82 & --
& $0.49^\circ[t_{1,2}]$ & 99.99 & 0.016
& \NoisePair{0.007}{0.025} & \NoisePair{0.0003}{0.049}
& \NoisePair{0.12}{0.16} & \NoisePair{0.105}{0.033}
& -- & 99.48 \\
\hline
5 & $\sqrt{\mathrm{iSWAP}}$ & 2
& 2.03 & 0.18 & 48.28 & 59.87 & 0.62 & 0.83 & --
& -$1.46^\circ[t_{1,2}]$ & 99.77 & 0.02
& \NoisePair{0.003}{0.07} & \NoisePair{0.01}{0.05}
& \NoisePair{0.002}{0.01} & \NoisePair{0.03}{0.007}
& -- & 99.56 \\
6 & $\sqrt{\mathrm{iSWAP}}$ & 150
& 0.27 & 0.32 & 140.8 & 127.3 & 1.02 & 1.38 & --
& -$39.4^\circ[t_{1,2}]$ & 99.62 & 0.005
& \NoisePair{0.032}{0.087} & \NoisePair{0.0028}{0.008}
& \NoisePair{0.009}{0.04} & \NoisePair{0.18}{0.005}
& -- & 99.25 \\
\hline
7 & CSWAP & 2
& 1.42 & 0.33 & 55.55 & 62.78 & 1.03 & 0.706 & -$0.614$
& \shortstack{-$3.9^\circ[c]$\\$70.3^\circ[t_{1,2}]$}
& 99.91 & 0.12
& \NoisePair{0.011}{0.123} & \NoisePair{0.0004}{0.01}
& \NoisePair{0.0006}{0.009} & \NoisePair{0.09}{0.04}
& \NoisePair{0.0016}{0.018} & 99.48 \\
8 & CSWAP & 150
& 0.4 & 0.38 & 134.4 & 205.75 & 1.75 & 1.23 & -$2.84$
& \shortstack{-$3.4^\circ[c]$\\$29.4^\circ[t_{1,2}]$}
& 99.79 & 0.03
& \NoisePair{0.006}{0.13} & \NoisePair{0.0009}{0.03}
& \NoisePair{0.0078}{0.032} & \NoisePair{0.18}{0.05}
& \NoisePair{0.03}{0.058} & 99.26 \\
\hline
9 & C-iSWAP & 2
& 1.75 & 0.19 & 49.18 & 94.55 & 1.44 & 0.81 & $-6.97$
& \shortstack{-$180^\circ[c]$\\$18.8^\circ[t_{1,2}]$}
& 99.996 & 0.15
& \NoisePair{0.0038}{0.03} & \NoisePair{0.00015}{0.044}
& \NoisePair{0.0009}{0.002} & \NoisePair{0.05}{0.028}
& \NoisePair{0.00014}{0.026} & 99.65 \\
10 & C-iSWAP & 150
& 1.47 & 0.15 & 55.17 & 117.05 & 1.26 & 0.71 & 20.07
& \shortstack{-$175.3^\circ[c]$\\-$74^\circ[t_{1,2}]$}
& 99.97 & 0.12
& \NoisePair{0.004}{0.03} & \NoisePair{0.0005}{0.117}
& \NoisePair{0.057}{0.05} & \NoisePair{0.09}{0.03}
& \NoisePair{0.000066}{0.014} & 99.46 \\
\hline
\end{tabular}%
}
\caption{Optimized control parameters, error budgets, and fidelities for
the SWAP-family gates. For each gate and temperature, the gate
parameters were optimized independently, accounting for fluctuations in
laser intensity and Rydberg interaction strengths, Doppler-induced
detuning fluctuations, and Rydberg-state decay. The drive is an
endpoint-subtracted Gaussian,
$\Omega_1(t)=\Omega_{1,\max}
\bigl[e^{-(t-T/2)^2/(2\sigma^2)}-e^{-T^2/(8\sigma^2)}\bigr]/
\bigl[1-e^{-T^2/(8\sigma^2)}\bigr]$ for $0\leq t\leq T$.
The reported parameters are rounded. The nominal coherent fidelity
$F_{\mathrm{coh}}$ is the average gate fidelity of
Eq.~\eqref{FidDefenition}, evaluated without fluctuations or decay and
after applying the listed single-qubit phase corrections to the
control ($c$) and targets ($t_{1,2}$) as indicated.
Each noise entry gives the excess population-rotation infidelity
(top) and relative-phase infidelity (bottom) for one noise source.
The fractional intensity and interaction fluctuations are zero-mean
quasistatic Gaussian variables with RMS widths of $1\%$ for
$\delta I_1/I_1$ ($0.2\%$ in row~1), $0.2\%$ for $\delta I_2/I_2$,
and $1\%$ for each fractional interaction fluctuation. For controlled
gates, the same fractional $V_{ct}$ fluctuation is applied to both
control--target pairs. The Doppler calculation uses the temperature
listed for each row. The total fidelity $F_{\mathrm{tot}}$ includes
all applicable fluctuations and Rydberg decay simultaneously.
Rydberg loss is averaged over the computational-basis inputs using
$\Gamma_r=1.6\,\mathrm{ms}^{-1}$. The lower Rydberg loss at
$150\,\mu\mathrm{K}$ in some gates is associated with their shorter optimized pulses. The gate outputs
are listed in Tables~S1 and S2.}
\label{tab_1}
\label{tab_2}
\end{table*}

Examining the pathways of  SWAP operation $\ket{01}\bra{10}$ in Fig.~\ref{Fig1}(c), we observe that the second-order transitions via $\ket{00}$ and $\ket{11}$, with amplitudes $-\Omega_1^2/4\Delta$ and $\Omega_1^2/4\Delta$ respectively, exactly cancel each other. 
This cancellation aligns with expectations since these paths do not involve any interacting states.
This cancellation allows the microwave drive $\Omega_1(t)$ to be
applied globally, while the optical field $\Omega_2$ selects the
participating targets. On spectator atoms outside the $\Omega_2$
beams, the leading effect of the far-detuned microwave drive is a
differential AC Stark shift of the qubit transition. The resulting
relative phase between $\ket{0}$ and $\ket{1}$ can be compensated
by a single-qubit $Z$ rotation or tracked in the qubit reference frame.

  Applying $\Omega_2$ to the targets opens the collective fourth-order pathway in Fig.~1(d), thereby enabling their exchange.
The Rydberg blockade forms the symmetric collective state ${(\ket{1r}+\ket{r1})/\sqrt{2}}$, which enables the SWAP operation through the coherent four-photon transition sequence
\begin{equation}\label{Eq_4photon}
\ket{10}\xrightarrow{\Omega_1} \ket{11}\xrightarrow{\sqrt{2}\Omega_2} \ket{\overline{1r}}\xrightarrow{\sqrt{2}\Omega_2} \ket{11}\xrightarrow{\Omega_1} \ket{01},
\end{equation}
as shown in Fig.~\ref{Fig1}d. 
If either target qubit is blocked by the Rydberg control state, the transition pathway through an individual excitation, such as $\ket{1r}$ or $\ket{r1}$, becomes inaccessible, and the SWAP process is consequently blocked, see Fig.~\ref{Fig1}e and the dashed lines in Fig.~\ref{Fig2}(b).

In the desired operation regime   $|V_{tt}+\delta_r-\Delta|\gg|\Omega_2|$,
$|\Delta-\delta_r|\gg|\Omega_2|$, $\Delta\gg(\Omega_1,\Omega_2)$, we
adiabatically eliminate the off-resonant states in
Fig.~\ref{Fig2}(a) and (d). The
dynamics in $\{\ket{01},\ket{10},\ket{\overline{1r}}\}$ is
described by
\begin{equation}\label{Eq_2}
\begin{aligned}
H={}&A[(\ket{01}+\ket{10})\bra{\overline{1r}}
+\mathrm{h.c.}]\\
&+B\ket{\overline{1r}}\bra{\overline{1r}}
+C(\ket{01}\bra{01}+\ket{10}\bra{10}),
\end{aligned}
\end{equation}
where
\begin{equation}\label{Eq_ABC}
\begin{aligned}
A&=\frac{\sqrt{2}\Omega_1\Omega_2}{4\Delta},\quad
B=\delta_r-\frac{\Omega_2^2}{2\Delta}
-\frac{\Omega_2^2}{2(V_{tt}+\delta_r-\Delta)},\\
C&=\frac{\Omega_2^2}{4(\Delta-\delta_r)}.
\end{aligned}
\end{equation}
The last term in $B$ is the second-order light shift from virtual excitation of $\ket{rr}$, which couples to
$\ket{\overline{1r}}$ with matrix element $\Omega_2/\sqrt{2}$. In the regime $|B-C|\gg\sqrt{2}|A|$, the effective
population Rabi frequency for exchange between
$\ket{01}$ and $\ket{10}$ is
\begin{equation}\label{eq:effective-exchange}
\Omega_{\mathrm{eff}}
\simeq \frac{2A^2}{|B-C|}.
\end{equation}

The physics of the exchange dynamics and accompanying phases could also be explained by the 
interference between the eigenstate contributions of
Eq.~\eqref{Eq_2}, which are
\begin{equation}\label{eq:dressed-states}
\begin{aligned}
\ket{\beta_0}&=\frac{\ket{01}-\ket{10}}{\sqrt{2}},\\
\ket{\beta_\pm}&=\mathcal{N}_\pm\bigl[
A(\ket{01}+\ket{10})+(\lambda_\pm-C)\ket{\overline{1r}}\bigr],
\end{aligned}
\end{equation}
 $\mathcal{N}_\pm$ are normalization factors and the eigenvalues are
\begin{equation}\label{eq:dressed-energies}
\lambda_0=C,\qquad
\lambda_\pm=\frac{B+C\pm\sqrt{(B-C)^2+8A^2}}{2}.
\end{equation}
During the target pulse, $B$ and $C$ are constant, while
$A(t)$ follows $\Omega_1(t)$. Let $\ket{\beta_g}$ denote the
dressed branch connected to $(\ket{01}+\ket{10})/\sqrt{2}$
as $A\to0$, its eigenvalue is $\lambda_g=\lambda_-$ for
$B>C$ and $\lambda_g=\lambda_+$ for $B<C$.

For a pulse with $A(0)=A(T)=0$, the initial state
$\ket{01}$ is an equal superposition of $\ket{\beta_g(0)}$
and $\ket{\beta_0}$.
Under adiabatic evolution,
the state is
\begin{equation}\label{Eq_exchange_state}
\begin{aligned}
&\ket{\psi(t)}
\simeq\frac{e^{-i\bar\phi(t)}}{\sqrt{2}}
\bigl[e^{-i\theta(t)}\ket{\beta_g(t)}
+e^{i\theta(t)}\ket{\beta_0}\bigr]\\
&=e^{-i\bar\phi(t)}\Bigl\{
\frac{\mathcal{M}_g(t)e^{-i\theta(t)}+e^{i\theta(t)}}{2}\ket{01}\\
&\quad+\frac{\mathcal{M}_g(t)e^{-i\theta(t)}-e^{i\theta(t)}}{2}\ket{10}-\frac{\mathcal{M}_g(t)A(t)e^{-i\theta(t)}}{B-\lambda_g(t)}
\ket{\overline{1r}}\Bigr\},
\end{aligned}
\end{equation}
where
\begin{equation}\label{eq:dressed-normalization}
\mathcal{M}_g(t)=
\left[1+\frac{2A(t)^2}{[B-\lambda_g(t)]^2}\right]^{-1/2}
\end{equation}
normalizes the dressed branch, and
\begin{equation}\label{eq:exchange-phases}
\begin{aligned}
\bar\phi(t)&=\frac{1}{2}\int_0^t
[\lambda_g(t')+\lambda_0]\,\mathrm{d}t',\\
\theta(t)&=\frac{1}{2}\int_0^t
[\lambda_g(t')-\lambda_0]\,\mathrm{d}t'.
\end{aligned}
\end{equation}
The two eigenstate contributions add in the amplitude of
$\ket{01}$ and subtract in that of $\ket{10}$, making their
relative phase $2\theta$ responsible for the exchange.
The common dynamical phase is $-\bar\phi$.
The Rydberg amplitude is explicitly proportional to $A(t)$.
At the pulse endpoints, $A=0$ and $\mathcal{M}_g=1$, so this
population vanishes in the adiabatic limit and the computational
amplitudes reduce to
\begin{equation}\label{eq:final-state}
 \ket{\psi(T)}=e^{-i\bar\phi(T)}(\cos\theta(T) \ket{01}-i\sin\theta(T)\ket{10}).
 \end{equation}
For the active input $\ket{1_c01}$, $\delta_r=0$. The accumulated eigenvalue splitting drives the exchange shown
in Fig.~\ref{Fig2}(b). Within the adiabatic description, the first complete
population transfer occurs when
\begin{equation}\label{Eq_T}
\left|\int_0^T
[\lambda_g(t)-\lambda_0]\,\mathrm{d}t\right|
=2|\theta(T)|=\pi.
\end{equation}
For a splitting that retains its sign throughout the pulse,
this is equivalent to
$\int_0^T\Omega_{\mathrm{eff},1_c}(t)\,\mathrm{d}t=\pi$.
At the same time, the pulse dependence of $\lambda_g(t)$
changes the phase-accumulation rate
$\dot{\bar\phi}=[\lambda_g(t)+C]/2$, accounting for the
curved portions of the phase evolution in
Fig.~\ref{Fig2}(c).

For the blocked input $\ket{r_c01}$, $\delta_r=V_{ct}$.
Away from the intermediate-state resonances, when
$|V_{ct}|$ dominates the light-shift corrections,
$B-C\simeq V_{ct}$ and the exchange rate is suppressed:
\begin{equation}\label{eq:blockade-ratio}
\frac{\Omega_{\mathrm{eff},r_c}}
     {\Omega_{\mathrm{eff},1_c}}
\simeq
\frac{\Omega_2^2}{4|V_{ct}\Delta|}
\left|\frac{3V_{tt}-\Delta}{V_{tt}-\Delta}\right|
\ll1.
\end{equation}
This estimate assumes that both branches satisfy the
elimination conditions above. Over the chosen gate duration,
$|\theta_{r_c}(T)|\ll1$ then ensures that the two eigenstate
contributions to $\ket{10}$ nearly cancel, blocking the
SWAP as in Fig.~\ref{Fig2}(e). Since
$\lambda_g\simeq\lambda_0=C$, the common phase accumulation
is approximately
\begin{equation}\label{eq:blocked-phase}
\bar\phi_{r_c}(t)\simeq Ct
=\frac{\Omega_2^2}{4(\Delta-V_{ct})}\,t.
\end{equation}
This phase is linear in time, as in Fig.~\ref{Fig2}(f).
The finite-$V_{tt}$ shift enters both the exchange and phase
evolution through $B$; it preserves this interference
picture as long as the relevant detunings remain large.

   \begin{figure*}
\centering
\includegraphics[width=\linewidth]{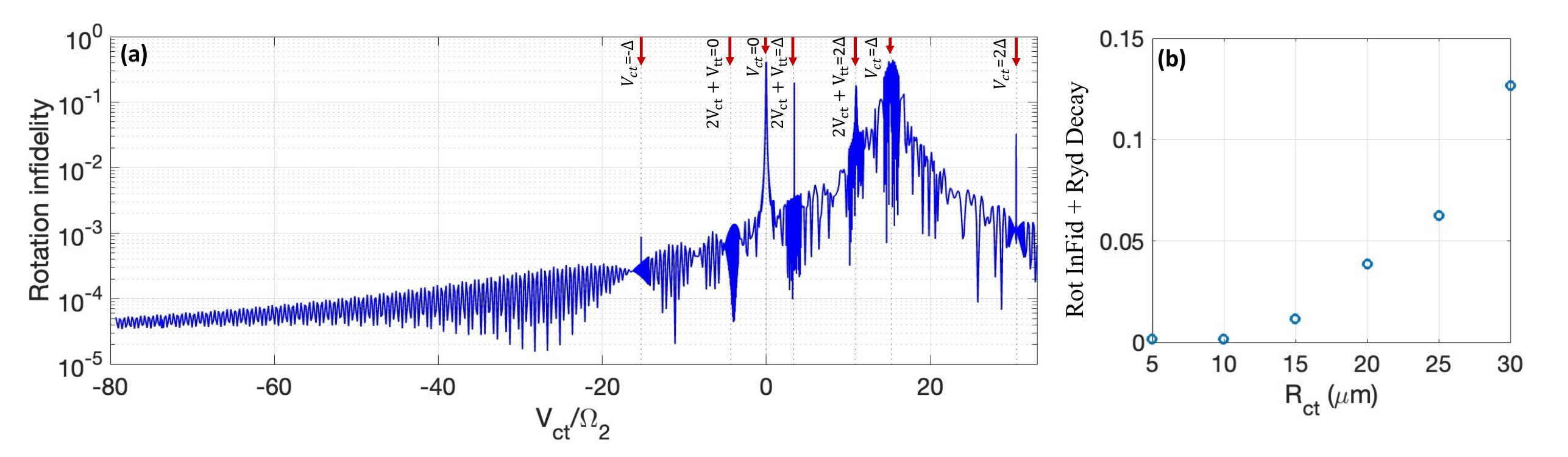}
\caption{\textbf{Long-range controlled-iSWAP operation.} \textbf{(a)} Rotation infidelity as a function of the normalized control–target interaction $V_{ct}/\Omega_2$, with all other parameters fixed to the values in row 9 of Table~I. The rotation infidelity remains low for attractive interactions as weak as $V_{ct}\simeq-1.5\Omega_2$ and rises sharply near $V_{ct}=0$, where the control can no longer block the exchange. The annotated conditions mark bare energy degeneracies, including those involving the doubly excited target state at $2V_{ct}+V_{tt}=0,\Delta,$ and $2\Delta$. \textbf{(b)} Combined error from imperfect population rotation and Rydberg-state decay as a function of the control–target separation $R_{ct}$ in a triangular geometry. The gate parameters are optimized independently at each separation for the Rydberg state $\lvert100P_{3/2},m_J=3/2\rangle$, with $C_6=-80\,\mathrm{THz}\,\mu\mathrm{m}^{6}$. 
Relative-phase errors are excluded in both panels; see Ref.~\cite{Explain} for the definition of rotation fidelity.}\label{Fig3}
\end{figure*}

For the input $\ket{1_c00}$, the pathways in
Fig.~\ref{Fig2}(g) connect $\ket{00}$ to $\ket{0r}$ and
$\ket{r0}$ through the off-resonant intermediate states
$\ket{01}$ and $\ket{10}$. A useful reference is the leading
resonant-pathway Hamiltonian, retaining these intermediate
states,
\begin{equation}\label{eq:dark-hamiltonian}
\begin{aligned}
H={}&\frac{\Omega_1}{2}(\ket{01}+\ket{10})\bra{00}
+\frac{\Omega_2}{2}
(\ket{0r}\bra{01}+\ket{r0}\bra{10})+\mathrm{h.c.}\\
&+\Delta'(\ket{01}\bra{01}+\ket{10}\bra{10}),
\end{aligned}
\end{equation}
where $\Delta'$ includes the leading light-shift correction
to their detuning.
This reduced Hamiltonian has the dark eigenstate
\begin{equation}\label{Eq_dark}
\ket{D(t)}=
\frac{\Omega_2\ket{00}
-\Omega_1(t)(\ket{0r}+\ket{r0})}
{\sqrt{\Omega_2^2+2\Omega_1(t)^2}},
\end{equation}
with zero eigenenergy. In the ideal adiabatic limit, its
individual Rydberg-state populations are $\Omega_1(t)^2/(\Omega_2^2+2\Omega_1(t)^2)$,
 For real drives and a pulse starting
and ending at $\Omega_1=0$, perfect following of this state
would return the system to $\ket{00}$ without a dynamical
phase.
For the pulse parameters used in Fig.~\ref{Fig2}, the
evolution departs from perfect dark-state following.
The time-dependent mixing angle,
$\alpha(t)=\arctan[\sqrt{2}\Omega_1(t)/\Omega_2]$, introduces
nonadiabatic coupling to bright  states when it varies
too rapidly compared with the relevant energy gaps.
The resulting bright-state admixture modifies the transient
populations and can generate a nonzero phase even when the
final $\ket{00}$ population is nearly restored.

In contrast, applying the target pulses to the initial state $\ket{r}_c\ket{00}_t$, with the level structure of Fig.~2(j), the control-target Rydberg interaction far-detunes all transitions, thus breaking the resonance and suppressing any population transfer, as illustrated in Fig.~\ref{Fig2}(k). 
Nonetheless, a dynamical phase is accumulated as depicted in Fig.~\ref{Fig2}(l). The dominant light shift comes from the second-order couplings of $\ket{00}$ with $\ket{01}$ and $\ket{10}$, adding up to $2\times\frac{\Omega_1^2(t)}{4\Delta}$. 
In qubit configurations $\ket{r_c}\ket{11}$ and $\ket{1_c}\ket{11}$, all the transitions are far detuned and hence the population transfer is minor.
In case of $\ket{r_c}\ket{11}$ the dominant light-shift  $\Delta E_{r_c11}=2\frac{\Omega_1^2}{4\Delta}+\frac{\Omega_2^2}{2(\Delta-V_{ct})}$ comes from the second order detuned coupling of $\ket{11}$ with $\ket{ \overline{1r}}$, $\ket{10}$ and $\ket{01}$, see Fig.~\ref{Fig2}(p).  
In the case of $\ket{1_c}\ket{11}$ with the level scheme of Fig.~\ref{Fig2}(m), a similar level scheme with the absence of control-target interaction $V_{ct}=0$ leads to the phase $\phi_{1_c11}=\frac{\Omega_1^2+\Omega_2^2}{2\Delta}T$.

Finally, the presented scheme can realize multiple SWAP‐type gates by omitting the control atom and tuning the pulse parameters listed in Table~\ref {tab_1}. Final phase adjustment can be applied with  $R(\phi)=\ket{0}\bra{0}+\exp(i\phi) \ket{1}\bra{1}$, and the so-called bSWAP operation can be realized as bSWAP=($\mathbb{I}$X)(SWAP)($\mathbb{I}$X).
The resulting error budgets for population rotation, phase adjustment, and Rydberg-state decay of all variants are compiled for each input qubit configuration in Tables~S1 and S2.

\subsection*{Gate fidelity, interaction range and laser power}
We quantify gate performance using the average gate fidelity \cite{Mol07}, 
\begin{equation}
\label{FidDefenition}
F=[\text{Tr}(MM^{\dagger})+|\text{Tr}(M)|^2]/[n(n+1)]
\end{equation}
with $M=U_{id}^{\dagger}U_{gate}$, where $U_{id}$ and $U_{gate}$, represents ideal and realistic gate operations.  $U_{gate}$ is obtained from numerical simulation of the gate for all the possible 8-qubit product state configurations.

In addition to imperfections in rotation and phase adjustment, the {\it short lifetime of Rydberg states} is a significant source of decoherence in atomic processors. As shown in Table~\ref{tab_2}, controlled-SWAP gates are more sensitive than SWAP gates to this decoherence mechanism because the control atom remains in the Rydberg state throughout the gate operation. Shorter gate durations are therefore particularly beneficial for controlled-SWAP gates. 
 To compare the presented blockade model with the conventional circuit model implemented with Rydberg gates, we define the time-integrated probability of being in the Rydberg state $\bar{T}_r= \int P_r(t) dt$, where $P_r(t)$ is the Rydberg population averaged over all qubit configurations at time $t$. For the parameter set of Fig.~\ref{Fig2}, the time-integrated Rydberg population over the gate operation is $\bar{T}_r=0.93\mu$s. 
In an alternative circuit model, implementing the C-iSWAP gate requires eight C-NOT gates. Applying the Gaussian $\pi$ pulse with the same laser intensity as in Fig.~\ref{Fig2}, to make the eight C-NOTs, the total time-integrated Rydberg population would be $\bar{T}_r=8.8\mu$s. Hence, the proposed scheme suppresses the time-integrated Rydberg population by an order of magnitude.

To assess the feasibility of controlled operations between widely separated qubits, we examine the dependence of the C-iSWAP gate on the control–target interaction strength and separation. Figure~\ref{Fig3}(a) shows the rotation infidelity as $V_{ct}$ is varied, with the remaining parameters fixed to the values in row~9 of Table~I. Low rotation errors persist for attractive interactions as weak as $V_{ct}\simeq-1.5\,\Omega_2$. Near $V_{ct}=0$, the control-induced blockade becomes insufficient to suppress the exchange between the target qubits, leading to a pronounced increase in infidelity.

Additional error features arise near interaction-induced resonances. The annotations in Fig.~\ref{Fig3}(a) identify the bare degeneracy conditions $V_{ct}=-\Delta,0,\Delta,2\Delta$ and $2V_{ct}+V_{tt}=0,\Delta,2\Delta$. The latter correspond to degeneracies between the doubly excited target state $\lvert rr\rangle$ and the computational states $\lvert00\rangle$, $\lvert01\rangle$ or $\lvert10\rangle$, and $\lvert11\rangle$, respectively, while the control atom occupies its Rydberg state. The resulting error features depend on the coupling strengths and pulse dynamics, and light shifts can displace their maxima from these bare resonance conditions.

Figure~\ref{Fig3}(b) evaluates the spatial range of the C-iSWAP operation in a triangular atomic geometry. It shows the combined error from imperfect population rotation and Rydberg-state decay as a function of the control–target separation $R_{ct}$, with the control parameters optimized independently at each separation. We use the Rydberg state $\lvert100P_{3/2},m_J=3/2\rangle$, with an attractive interaction coefficient $C_6=-80\,\mathrm{THz}\,\mu\mathrm{m}^{6}$. To reduce the computational cost of these scans, relative-phase errors are excluded from the optimization and the plotted error measures in both panels of Fig.~3. An alternative configuration uses $\lvert70S_{1/2},m_J=1/2\rangle$ for the control and $\lvert67D_{5/2},m_J=5/2\rangle$ for the targets, providing a comparable attractive interaction strength while allowing two-photon excitation.

To estimate the laser power required for the couplings in Table~\ref{tab_1}, we consider two-photon Rydberg excitation in Cs through the $6S_{1/2}\rightarrow7P_{1/2}\rightarrow nS_{1/2}$ ladder using $459$-nm and approximately $1040$-nm fields, as employed in neutral-atom arrays \cite{deOliveira2025}. For super-Gaussian intensity profiles $I(r)=I_0\exp[-(r/\sigma)^{10}]$ with a $3\mu\mathrm{m}$ $1/e$-intensity diameter, an intermediate-state detuning of $|\Delta_p|/2\pi=10$ GHz, and equal single-photon Rabi frequencies, we estimate that $\Omega_2/2\pi=60$ - $235$ MHz requires approximately $5$ - $19.5$ mW at $459$ nm and $2.8$ - $10.8$ W in the infrared for $\ket{100S}$ excitation. For $70S$, the stronger upper-leg matrix element lowers the infrared estimate to $0.9$ - $3.5\,\mathrm{W}$.
The relevant source-power scales are available, i.e., $340$ mW of tunable single-frequency light near $459$ nm has been demonstrated \cite{Jensen2013}, and a $20$W amplifier near $1038$ nm has been used in a Cs tweezer apparatus \cite{Wang2024}.

\subsection*{Robustness to experimental fluctuations}

\begin{figure*}
\centering
\includegraphics[width=\linewidth]{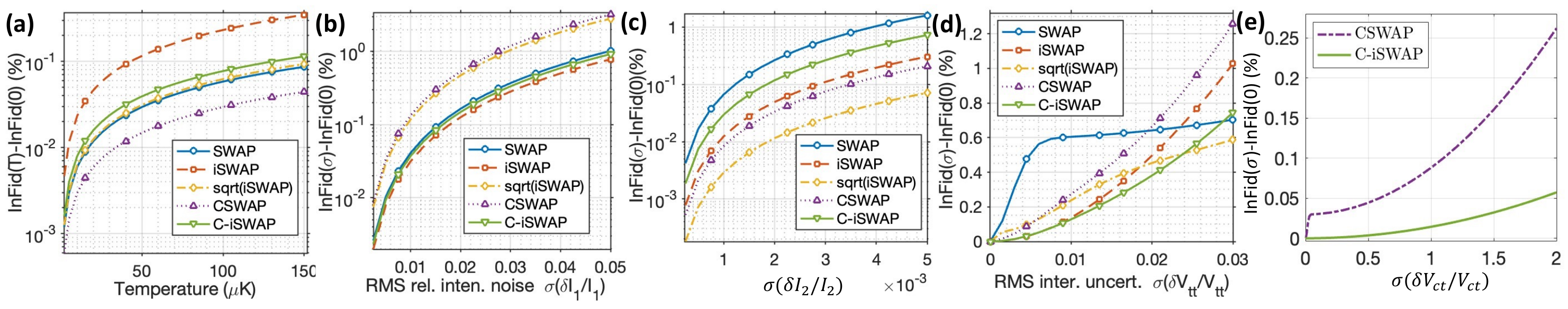}
\caption{{\bf Robustness of the five exchange gates.}
Excess average gate infidelity versus (a) atomic temperature, corresponding to Doppler shift,
(b,c) RMS fractional intensity fluctuations in $I_1$ and $I_2$,
(d) RMS fractional uncertainty in $V_{tt}$, and (e) RMS
fractional uncertainty in $V_{ct}$. The technical
fluctuations are zero-mean quasistatic Gaussian variables;
the control--target interaction applies only to the
controlled gates. Each curve subtracts the nominal
infidelity evaluated with the same Rydberg decay rate,
and the nominal single-atom phase corrections remain fixed.
The pulse parameters are the $150\,\mu\mathrm{K}$ optimized
sets in Table~\ref{tab_1}.}\label{Fig4}
\end{figure*}

Figure~\ref{Fig4} compares the excess infidelity of the five
exchange gates under experimental fluctuations. The operating
points are obtained by minimizing an objective function that combines
the nominal infidelity with the mean and dispersion of the
infidelity under simultaneous Doppler, intensity, and
interaction noise, with Rydberg decay included. The search
varies the detuning, both couplings, gate duration, pulse
width, and target--target interaction, as well as the
control--target interaction for controlled gates.
The single-atom phase corrections are calibrated at the
nominal operating point and held fixed for every noise
realization.
Table~\ref{tab_2} reports the decay and fluctuation error
budgets and total fidelities.

The Supplemental Material identifies the light shifts that
connect these fluctuations to relative-phase errors. In the
adiabatic limit, a quasistatic parameter variation produces
$\delta\phi_n=-\int_0^T\delta E_n(t)\,\mathrm{d}t$; the gate
error depends on differences between the phases of the
computational components. For equal control--target
interactions and constant $\Omega_2$, two representative
interaction-induced responses are
\begin{equation}\label{eq:interaction-phase-errors}
\begin{aligned}
\delta\phi_{r_c01}
&\simeq-\frac{T\Omega_2^2}{4(\Delta-V_{ct})^2}\,\delta V_{ct},\\
\delta\phi_{1_c11}^{(4)}
&\simeq-\frac{T\Omega_2^4}
{4\Delta^2(2\Delta-V_{tt})^2}\,\delta V_{tt}.
\end{aligned}
\end{equation}
The first follows from the second-order shift of a singly
excited target in the blocked branch; the second comes from
virtual double excitation in the active branch. Both require
the corresponding transitions to remain off resonance.
For $\ket{1_c01}$ input, $V_{tt}$ also changes the dressed
splitting responsible for exchange; see Eqs.~\eqref{Eq_exchange_state}--\eqref{Eq_T}. Hence, interaction noise
perturbs both the exchange angle and the output phases; see Supplementary Material.

Intensity fluctuations affect the same phase shifts. For a common
fractional fluctuation $\epsilon_2=\delta I_2/I_2$ of one
excitation field on both targets, with the other field and
the effective detuning fixed,
$\delta\Omega_2/\Omega_2\simeq\epsilon_2/2$.
The leading optical phase-response coefficients are
\begin{equation}\label{eq:intensity-coefficients}
a=-\frac{T\Omega_2^2}{4\Delta},\qquad
b=-\frac{T\Omega_2^2}{4(\Delta-V_{ct})}.
\end{equation}
In the target basis $(00,01,10,11)$, the residual phases are
$\epsilon_2(0,a,a,2a)$ in the active branch and
$\epsilon_2(0,b,b,2b)$ in the blocked branch. These random
relative phases remain after fixed nominal compensation.

These responses show how adjusting the relative couplings
can improve robustness. At fixed duration and detunings,
reducing $\Omega_2$ decreases the leading intensity-phase
response as $\Omega_2^2$, the $V_{ct}$ phase response as
$\Omega_2^2$, and the displayed $V_{tt}$ response as
$\Omega_2^4$. Within a dispersive pulse family of fixed shape and
interaction-to-detuning ratios, away from resonances of Fig.~3a,
maintaining the exchange
angle approximately fixes $T\Omega_{1,\max}^2/|\Delta|$,
so $|a|$ scales as
$(\Omega_2/\Omega_{1,\max})^2$. However, reducing this ratio
also changes the dressed-state gaps and can increase
nonadiabatic errors. Increasing $\Delta$ reduces
instantaneous shifts but can lengthen the gate, leaving
$T\Omega_2^2/\Delta$ approximately unchanged. The
$\Omega_1$ drive also controls the exchange area and remains
sensitive to $I_1$ noise; see  Supplementary Material for detailed discussions.  The optimization balances
coupling ratios, detunings, and pulse duration to reduce
relative-phase and exchange-angle errors while preserving
population return and limiting Rydberg decay. The objective
penalizes the resulting noisy infidelity directly.
The optimized durations are retained in these estimates,
and the final error budget is evaluated numerically in Table~\ref{tab_2}.

We represent the Doppler effect by assuming Gaussian-distributed velocities and hence detunings  with 
$\sigma_\delta=k_{\mathrm{eff}}\sqrt{k_BT_a/M}$ for
counter-propagating 1040-nm and 459.6-nm fields in Cs.
The technical-noise sweeps assume zero-mean quasistatic Gaussian
intensity and interaction fluctuations. The plotted quantity
is the noise-averaged infidelity minus its nominal value,
with the same Rydberg decay rate and fixed phase corrections
in both terms. Its initial quadratic dependence on RMS
noise follows from averaging a smooth response over
zero-mean fluctuations; the nearly linear temperature
dependence follows from the Doppler variance being
proportional to $T_a$.

At $T_a=150\,\mu\mathrm{K}$, the excess Doppler infidelity
ranges from $0.044\%$ for CSWAP to $0.348\%$ for iSWAP
[Fig.~\ref{Fig4}(a)]. At $1\%$ RMS fluctuations in $I_1$,
all five gate infidelities remain below $0.135\%$
[Fig.~\ref{Fig4}(b)]. At $0.2\%$ RMS fluctuations in $I_2$,
the excess infidelity ranges from $0.011\%$ for
$\sqrt{\mathrm{iSWAP}}$ to $0.266\%$ for SWAP
[Fig.~\ref{Fig4}(c)]. A $1\%$ RMS uncertainty in $V_{tt}$
gives $0.128$--$0.605\%$ across the five gates
[Fig.~\ref{Fig4}(d)], while the same fractional uncertainty
in $V_{ct}$ gives $0.088\%$ for CSWAP and $0.014\%$ for
C-iSWAP [Fig.~\ref{Fig4}(e)]. Thus, the dominant error source
depends on the gate and operating point; blockade suppresses
unwanted population exchange but does not remove phase
sensitivity.

Optimizing the parameters of a single Gaussian pulse already reduces sensitivity to laser-intensity fluctuations. Further suppression could be explored through optimal control, which can search for the refocusing used in composite pulses. An adiabatic echo emerged from such a search for robust many-body state preparation \cite{Zen26}. For a controlled-exchange gate, the same pulse must produce the desired exchange on the active branch while preserving its phase relative to the blocked branch.

Experimental control complements this parameter optimization.
Super-Gaussian beams with flat intensity profiles can reduce
spatial inhomogeneity \cite{Gil16}, while active stabilization
and shot-to-shot monitoring at the $0.2\%$ level have been
reported for Rydberg-excitation beams \cite{Lin2026}.
Microwave-induced transitions within the Rydberg manifold
must also be assessed for the chosen frequencies, field
amplitudes, polarizations, and spectra \cite{Eba21}. Where necessary,
an optical Raman implementation of $\Omega_1$ preserves
the effective coupling used by the exchange mechanism.

\subsection*{Comparison with anti-blockade controlled-exchange schemes}

Ref.~\cite{Wu21} proposes a controlled-exchange (Fredkin-type) operation in an anti-blockade setting, whereas our protocol operates in the standard Rydberg blockade regime. This distinction has direct consequences for experimental robustness.
 A  practical difference concerns shot-to-shot variations of the interatomic separation. In anti-blockade schemes, the distance-dependent interaction $V(d)$ enters the resonance condition and must therefore be known with high precision. For representative parameters used in Ref.~\cite{Wu21} (see p.~5 therein), even under tight confinement on the order of $\sim30$nm, the resulting interaction variation is comparable to the effective couplings that drive the near-resonant $\ket{01}\leftrightarrow\ket{rr}$ dynamics. Such interaction inhomogeneity translates directly into detuning errors and tightens calibration requirements. Related considerations apply to the detuned four-photon SWAP variant discussed in Ref.~\cite{Wu21}, where the effective exchange angle depends sensitively on the detuning from $\ket{rr}$. In other words, uncertainty in the interaction-induced shift can exceed the detuning scale, hence producing large fluctuations of the effective pulse area.

   \begin{figure*}
\centering
\includegraphics[width=\linewidth]{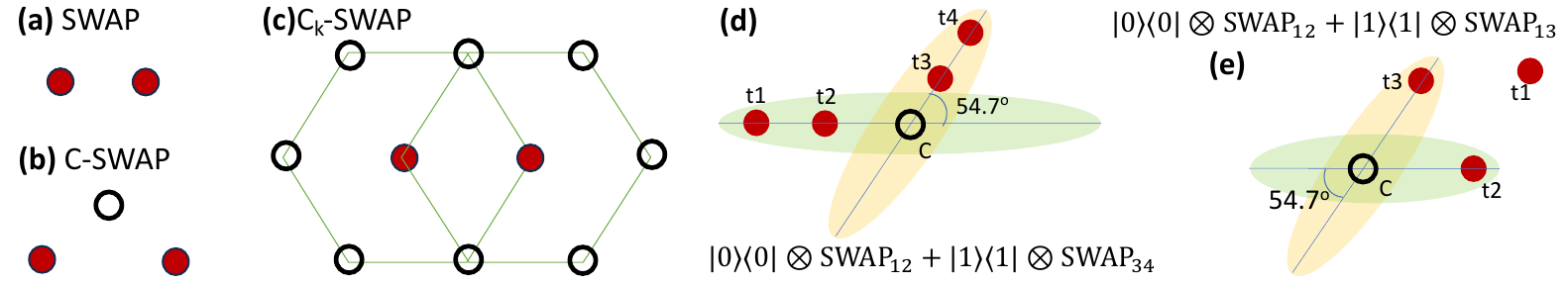}
\caption{{\bf Geometry of extended SWAP and C-SWAP variants with higher-order generalizations.}
Control and target qubits are represented by hollow and solid circles, respectively. Panels (a)–(c) illustrate the geometries of (a) a SWAP gate, (b) a controlled-SWAP (C-SWAP) gate, and (c) a multi-control SWAP (C$_k$-SWAP) gate,  arranged in a triangular lattice configuration in which every pair of atoms resides within the mutual Rydberg blockade range.
Panels (d,e) illustrate the geometries of conditional multiplexed SWAP gates. Different logical states of the control atom are coupled to distinct Rydberg levels, $|nP_{3/2},3/2\rangle$ or $|(n-1)D_{5/2},5/2\rangle$, while the targets are driven to $|nS_{1/2},1/2\rangle$. The resulting anisotropic control–target blockade regions, shown as green and yellow shaded areas, selectively inhibit the SWAP operation for blockaded targets.
(d) Four-target configuration implementing Eq.~\eqref{Eq_10}.
(e) Three-target architecture implementing Eq.~\eqref{Eq_11}, where target 1 always lies outside the control blockade, while targets 2 and 3 are blockaded conditionally on the control-qubit state.}\label{Fig5}
\end{figure*}

\section*{Discussion}

In this work, we have introduced and analyzed a general interference-blockade-enabled mechanism for realizing native exchange and controlled-exchange gates in neutral-atom quantum processors. By tuning a controllable near-degeneracy in the collective manifold of excited states, the scheme coherently toggles mediated exchange channels to perform single-step SWAP and controlled-SWAP (Fredkin) operations with fidelities exceeding 99.6\% and an order-of-magnitude reduction in circuit depth and Rydberg exposure compared with decomposed implementations. The protocol remains robust against Doppler broadening, realistic laser-intensity fluctuations, and interaction fluctuations, reducing thermal and calibration demands in large-scale arrays.

In the {\it Methods} section, we discuss that beyond single-control gates, the same interference principle enables (C$_k$–SWAP), where several control qubits jointly regulate SWAP operation. We also discuss conditional multiplexed SWAP extensions, where a control qubit regulates or selects among multiple exchange channels. These multi-path and multi-flag generalizations provide compact hardware primitives for conditional routing, multi-copy verification, and syndrome-conditioned logic, advancing neutral-atom systems toward coherent quantum routers and QRAM elements.

By elevating exchange to a native primitive alongside conditional phase operations, this work establishes a versatile tool set for neutral-atom computation. Our work may inspire the pursuit of similar programmable multiqubit exchange in other architectures and mark a critical step toward scalable verification, efficient fermionic simulation, and hardware-level compilation across the broader landscape of quantum technologies.

\section*{Methods}
\subsection*{Multi-control SWAP gates}
The scheme can be extended beyond the single-control Fredkin gate to realize a multi-control SWAP (C$_k$–SWAP), where $k$ control qubits jointly regulate the exchange of two target registers.
Considering the geometry of Fig.~\ref{Fig5}c, any two atoms lie within the mutual blockade range. Control qubits are sequentially excited from the $\ket{0}$ logic state to the Rydberg state before applying the target pulses of Fig.~\ref{Fig1} and are de-excited in reverse order.
 In this configuration, the exchange channel becomes resonant only when all control qubits remain in the $\ket{1}$ state, while any Rydberg excitation among them induces strong control–target interactions that block the SWAP. As shown in Fig.~\ref{Fig3}(a), for attractive control–target couplings, interaction strengths exceeding about $3\Omega_2$ are sufficient to block the exchange, ensuring that only the fully unexcited control configuration participates in the resonant transfer. 
This provides a compact and hardware-efficient realization of multi-flag conditional logic, directly applicable to syndrome-conditioned operations in quantum error correction, multi-copy state-verification protocols (e.g., multi-register SWAP tests), and heralded entanglement routing where several control nodes jointly authorize an exchange event.

\subsection*{Conditional multiplexed SWAP gates}
A complementary generalization is the {\it conditional multiplexed SWAP}, defined as
\begin{equation}
U = |0\rangle_c\langle 0| \otimes \mathrm{SWAP}_{12} + |1\rangle_c\langle 1| \otimes \mathrm{SWAP}_{34},
\label{Eq_10}
\end{equation}
where the control qubit determines which target pair undergoes exchange; see Fig.~\ref{Fig5}d.
This operation realizes a conditional multi-path exchange, where the control qubit selects between two disjoint exchange channels. It functions as a quantum switchboard, coherently activating one of several communication links between independent register pairs. In superposition, the control can coherently engage multiple exchange pathways simultaneously, allowing interference-based routing and parallel quantum-state transfer. Such functionality provides a compact primitive for reconfigurable quantum networks and entanglement-routing protocols.

Using anisotropic Rydberg interactions, the control and target qubits can be arranged so that distinct target pairs couple through different Rydberg manifolds. The control qubit is excited via a $\pi$ pulse from different logic qubits to either $\ket{nP_{3/2},3/2}$ or $\ket{(n-1)D_{5/2},5/2}$, while the four target atoms are driven by identical pulses of Fig.~\ref{Fig1} in addressing the $\ket{nS_{1/2},1/2}$ Rydberg state.
Exploiting the anisotropy of the nS–nP and nS–nD interactions, one target pair is aligned along the quantization axis, and the other is positioned at the magic angle $\theta=54.7^\circ$, where their interaction contrast is maximized. This configuration renders the control–target couplings spin-dependent, enabling selective blockade of one target pair while leaving the other dark to the control interaction; see Fig.~\ref{Fig5}d.

As discussed, in the four-photon SWAP process of Eq.~\eqref{Eq_4photon}, the exchange occurs only through the symmetric collective state $(\ket{1r}+\ket{r1})/\sqrt{2}$. If either target qubit is blockaded by the Rydberg control, the intermediate single-excitation states $\ket{1r}$ or $\ket{r1}$ cannot support the SWAP transition. In the four-target configuration, the control qubit thus blocks Rydberg excitation in one target pair while the other, free from control-induced interaction, would exclusively undergo the SWAP operation.

In a different three-target architecture of Fig.~\ref{Fig5}e, by arranging three target atoms such that target 1 always lies outside the control’s blockade region, while targets 2 and 3 are blockaded conditionally on the control qubit state, one can realize
\begin{equation}
U = |0\rangle_c\langle 0| \otimes \mathrm{SWAP}_{12} + |1\rangle_c\langle 1| \otimes \mathrm{SWAP}_{13},
\label{Eq_11}
\end{equation}
where the control qubit determines which secondary register (2 or 3) would exchange its quantum state with a shared site 1. Note that only the pair of target atoms that are not Rydberg-blocked by the control, but lie within each other’s blockade radius, can participate in the four-photon SWAP process via the symmetric collective state $(\ket{1r}+\ket{r1})/\sqrt{2}$.
Unlike the four-target design, which connects two independent pairs, this geometry introduces a common hub qubit that coherently couples to multiple destinations.
Such a configuration acts as a coherent quantum memory port, enabling controlled transfer between two memory registers and a shared central node.
In a networked setting, it provides quantum-controlled link selection, where the control qubit decides which remote site establishes a connection with the central hub.
Because the process is fully coherent, superpositions of control states realize simultaneous and interference-based routing, establishing a hardware primitive for quantum random-access memory \cite{Giovannetti2008} and dynamically reconfigurable entanglement networks.\\

\section*{Author Contributions}
MK developed the gate protocols and performed all calculations. Both authors, MK and KM, contributed equally to the analysis of the results and to the writing of the manuscript.

\section*{Competing Interests}
The authors declare no competing financial or non-financial interests.

\section*{Data Availability}
The MATLAB scripts and reference outputs for the coherent gate fidelities and input-resolved Rydberg loss probabilities are available on Figshare at https://doi.org/10.6084/m9.figshare.35117536\\
Given the large number of simulations, the remaining simulation outputs underlying the figures and tables are available from the corresponding author upon reasonable request.
\section*{Acknowledgements}
Not Applicable.

\section*{Funding}
K. M\o lmer acknowledges support from the Danish National Research Foundation Centre of Excellence “Hy-Q,” Grant No. DNRF139 and the Carlsberg Foundation Semper Ardens project QCooL.
M. Khazali did not receive funding for this work.

\clearpage
\onecolumngrid
\setcounter{section}{0}
\setcounter{table}{0}
\setcounter{figure}{0}
\setcounter{equation}{0}
\renewcommand{\thetable}{S\arabic{table}}
\renewcommand{\thefigure}{S\arabic{figure}}
\renewcommand{\theequation}{S\arabic{equation}}

\section*{ Supplemental Materials for ``Blockade-induced exchange primitives for scalable neutral-atom QPU"}


\begin{table*}[h!]
\centering
\begingroup
\scriptsize
\setlength{\tabcolsep}{3pt}
\renewcommand{\arraystretch}{1.15}
\resizebox{\textwidth}{!}{%
\begin{tabular}{|c|c|c|}
\hline
Gate & $U_{\mathrm{coh}}$ & $P_{\mathrm{RydLos}}$ \\
\hline
\rotatebox{90}{SWAP}
&
$\begin{pmatrix}
0.999987e^{-i0.0014} &
\textcolor{gray}{1.24{\times}10^{-4}e^{-i2.547}} &
\textcolor{gray}{1.24{\times}10^{-4}e^{-i2.547}} &
\textcolor{gray}{6.78{\times}10^{-9}e^{-i2.731}} \\[3pt]
\textcolor{gray}{1.24{\times}10^{-4}e^{i2.886}} &
\textcolor{gray}{0.00414e^{-i1.897}} &
0.998630e^{i0.0014} &
\textcolor{gray}{9.19{\times}10^{-4}e^{-i1.113}} \\[3pt]
\textcolor{gray}{1.24{\times}10^{-4}e^{i2.886}} &
0.998630e^{i0.0014} &
\textcolor{gray}{0.00414e^{-i1.897}} &
\textcolor{gray}{9.19{\times}10^{-4}e^{-i1.113}} \\[3pt]
\textcolor{gray}{6.78{\times}10^{-9}e^{i1.852}} &
\textcolor{gray}{9.19{\times}10^{-4}e^{-i1.963}} &
\textcolor{gray}{9.19{\times}10^{-4}e^{-i1.963}} &
0.999993e^{-i0.0014}
\end{pmatrix}$
&
$\operatorname{diag}\!\left(0.000665,\,0.000084,\,0.000084,\,0.000009\right)$
\\
\hline
\rotatebox{90}{iSWAP}
&
$\begin{pmatrix}
0.999999e^{-i0.0004} &
\textcolor{gray}{3.31{\times}10^{-5}e^{-i2.644}} &
\textcolor{gray}{3.31{\times}10^{-5}e^{-i2.644}} &
\textcolor{gray}{2.37{\times}10^{-10}e^{i1.673}} \\[3pt]
\textcolor{gray}{3.31{\times}10^{-5}e^{i1.149}} &
\textcolor{gray}{0.00247e^{-i3.135}} &
0.999978e^{i1.571} &
\textcolor{gray}{2.64{\times}10^{-4}e^{-i2.721}} \\[3pt]
\textcolor{gray}{3.31{\times}10^{-5}e^{i1.149}} &
0.999978e^{i1.571} &
\textcolor{gray}{0.00247e^{-i3.135}} &
\textcolor{gray}{2.64{\times}10^{-4}e^{-i2.721}} \\[3pt]
\textcolor{gray}{2.37{\times}10^{-10}e^{i2.975}} &
\textcolor{gray}{2.64{\times}10^{-4}e^{i1.072}} &
\textcolor{gray}{2.64{\times}10^{-4}e^{i1.072}} &
0.999992e^{-i0.0004}
\end{pmatrix}$
&
$\operatorname{diag}\!\left(0.000402,\,0.000123,\,0.000123,\,0.000012\right)$
\\
\hline
\rotatebox{90}{$\sqrt{\mathrm{iSWAP}}$}
&
$\begin{pmatrix}
0.999866e^{-i0.0045} &
\textcolor{gray}{5.52{\times}10^{-4}e^{i0.474}} &
\textcolor{gray}{5.52{\times}10^{-4}e^{i0.474}} &
\textcolor{gray}{5.40{\times}10^{-10}e^{-i1.922}} \\[3pt]
\textcolor{gray}{5.52{\times}10^{-4}e^{i1.021}} &
0.701736e^{i0.0027} &
0.709346e^{i1.577} &
\textcolor{gray}{4.08{\times}10^{-5}e^{-i1.417}} \\[3pt]
\textcolor{gray}{5.52{\times}10^{-4}e^{i1.021}} &
0.709346e^{i1.577} &
0.701736e^{i0.0027} &
\textcolor{gray}{4.08{\times}10^{-5}e^{-i1.417}} \\[3pt]
\textcolor{gray}{5.40{\times}10^{-10}e^{-i0.827}} &
\textcolor{gray}{4.08{\times}10^{-5}e^{-i0.869}} &
\textcolor{gray}{4.08{\times}10^{-5}e^{-i0.869}} &
0.999977e^{-i0.0045}
\end{pmatrix}$
&
$\operatorname{diag}\!\left(0.000776,\,0.000030,\,0.000030,\,0.000032\right)$
\\
\hline
\end{tabular}%
}
\endgroup
\caption{{\bf Numerical operations of the optimized two-target gates at $2\,\mu\mathrm{K}$.}
The phase-corrected coherent operators are given in the logical basis
$(|00\rangle,|01\rangle,|10\rangle,|11\rangle)$ using the strict
eight-state target model, including $|rr\rangle$ and omits the decoupled
antisymmetric Rydberg state $(|1r\rangle-|r1\rangle)/\sqrt{2}$.
(For the temperature sweeps in Fig.~4(a) and Table I of the main text, we retain nine states of main-text Eq.~(1) because
independent Doppler shifts on the two targets couple it to the symmetric Rydberg state.)
Matrix elements are written as $|U_{ij}|e^{i\phi_{ij}}$, with phases in radians; entries outside
the support of the ideal gate are shown in gray.  The physically irrelevant
global phase is subtracted so that the overlap with the ideal gate is real and
positive.  The $2\,\mu\mathrm{K}$ label specifies the optimized parameter
set; $U_{\mathrm{coh}}$ itself is the nominal zero-Doppler, zero-technical-noise
map with decay disabled.  For the same input ordering, the diagonal entries
of $P_{\mathrm{RydLos}}$ are the no-jump norm deficits obtained separately
with $\Gamma_r=1.6\times10^{-3}\,\mu\mathrm{s}^{-1}$.  Thus,
$P_{\mathrm{RydLos}}$ quantifies decay loss and does not include coherent
population remaining in noncomputational states.}
\label{tab:two-uncontrolled-maps}
\end{table*}

\begin{table*}[h!]
\centering
\begingroup
\scriptsize
\setlength{\tabcolsep}{3pt}
\renewcommand{\arraystretch}{1.15}
\resizebox{\textwidth}{!}{%
\begin{tabular}{|c|c|c|c|}
\hline
Gate & Branch & $U_{\mathrm{coh}}=B\oplus A$ & $P_{\mathrm{RydLos}}$ \\
\hline
\multirow{2}{*}{\rotatebox{90}{CSWAP}}
& $B$ &
$\begin{pmatrix}
1.000000e^{-i0.0038} &
\textcolor{gray}{2.90{\times}10^{-6}e^{i2.158}} &
\textcolor{gray}{2.90{\times}10^{-6}e^{i2.158}} &
\textcolor{gray}{8.69{\times}10^{-12}e^{-i1.863}} \\[3pt]
\textcolor{gray}{2.90{\times}10^{-6}e^{i0.985}} &
0.999835e^{i0.0023} &
\textcolor{gray}{0.00583e^{i1.574}} &
\textcolor{gray}{3.01{\times}10^{-6}e^{i2.159}} \\[3pt]
\textcolor{gray}{2.90{\times}10^{-6}e^{i0.985}} &
\textcolor{gray}{0.00583e^{i1.574}} &
0.999835e^{i0.0023} &
\textcolor{gray}{3.01{\times}10^{-6}e^{i2.159}} \\[3pt]
\textcolor{gray}{8.69{\times}10^{-12}e^{i2.074}} &
\textcolor{gray}{3.01{\times}10^{-6}e^{i0.987}} &
\textcolor{gray}{3.01{\times}10^{-6}e^{i0.987}} &
0.999609e^{-i0.0009}
\end{pmatrix}$
&
\multirow{2}{*}{%
$\operatorname{diag}\!\left(
\begin{array}{c}
0.002270,\,0.002272,\,0.002272,\,0.002273,\\
0.000843,\,0.000103,\,0.000103,\,0.000008
\end{array}\right)$}
\\
\cline{2-3}
& $A$ &
$\begin{pmatrix}
0.999964e^{i0.0116} &
\textcolor{gray}{1.80{\times}10^{-4}e^{-i0.500}} &
\textcolor{gray}{1.80{\times}10^{-4}e^{-i0.500}} &
\textcolor{gray}{1.05{\times}10^{-8}e^{-i0.553}} \\[3pt]
\textcolor{gray}{1.80{\times}10^{-4}e^{-i1.673}} &
\textcolor{gray}{0.02115e^{i1.546}} &
0.998712e^{-i0.0102} &
\textcolor{gray}{8.32{\times}10^{-4}e^{i2.205}} \\[3pt]
\textcolor{gray}{1.80{\times}10^{-4}e^{-i1.673}} &
0.998712e^{-i0.0102} &
\textcolor{gray}{0.02115e^{i1.546}} &
\textcolor{gray}{8.32{\times}10^{-4}e^{i2.205}} \\[3pt]
\textcolor{gray}{1.05{\times}10^{-8}e^{-i2.898}} &
\textcolor{gray}{8.32{\times}10^{-4}e^{i1.033}} &
\textcolor{gray}{8.32{\times}10^{-4}e^{i1.033}} &
0.999990e^{i0.0087}
\end{pmatrix}$
& \\
\hline
\multirow{2}{*}{\rotatebox{90}{C-iSWAP}}
& $B$ &
$\begin{pmatrix}
1.000000e^{-i0.0002} &
\textcolor{gray}{1.62{\times}10^{-8}e^{-i0.346}} &
\textcolor{gray}{1.62{\times}10^{-8}e^{-i0.346}} &
\textcolor{gray}{2.87{\times}10^{-16}e^{-i0.234}} \\[3pt]
\textcolor{gray}{1.62{\times}10^{-8}e^{-i2.795}} &
0.999979e^{i0.0001} &
\textcolor{gray}{1.15{\times}10^{-4}e^{i1.571}} &
\textcolor{gray}{1.59{\times}10^{-8}e^{-i0.346}} \\[3pt]
\textcolor{gray}{1.62{\times}10^{-8}e^{-i2.795}} &
\textcolor{gray}{1.15{\times}10^{-4}e^{i1.571}} &
0.999979e^{i0.0001} &
\textcolor{gray}{1.59{\times}10^{-8}e^{-i0.346}} \\[3pt]
\textcolor{gray}{2.87{\times}10^{-16}e^{i1.151}} &
\textcolor{gray}{1.59{\times}10^{-8}e^{-i2.795}} &
\textcolor{gray}{1.59{\times}10^{-8}e^{-i2.795}} &
0.999961e^{-i0.0001}
\end{pmatrix}$
&
\multirow{2}{*}{%
$\operatorname{diag}\!\left(
\begin{array}{c}
0.002803,\,0.002803,\,0.002803,\,0.002803,\\
0.000402,\,0.000123,\,0.000123,\,0.000012
\end{array}\right)$}
\\
\cline{2-3}
& $A$ &
$\begin{pmatrix}
0.999999e^{-i0.0005} &
\textcolor{gray}{3.06{\times}10^{-5}e^{-i2.667}} &
\textcolor{gray}{3.06{\times}10^{-5}e^{-i2.667}} &
\textcolor{gray}{2.42{\times}10^{-10}e^{i1.623}} \\[3pt]
\textcolor{gray}{3.06{\times}10^{-5}e^{i1.167}} &
\textcolor{gray}{0.00270e^{-i3.135}} &
0.999977e^{i1.571} &
\textcolor{gray}{2.68{\times}10^{-4}e^{-i2.758}} \\[3pt]
\textcolor{gray}{3.06{\times}10^{-5}e^{i1.167}} &
0.999977e^{i1.571} &
\textcolor{gray}{0.00270e^{-i3.135}} &
\textcolor{gray}{2.68{\times}10^{-4}e^{-i2.758}} \\[3pt]
\textcolor{gray}{2.42{\times}10^{-10}e^{i3.009}} &
\textcolor{gray}{2.68{\times}10^{-4}e^{i1.076}} &
\textcolor{gray}{2.68{\times}10^{-4}e^{i1.076}} &
0.999986e^{-i0.0007}
\end{pmatrix}$
& \\
\hline
\end{tabular}%
}
\endgroup
\caption{{\bf Numerical operations of the optimized controlled gates at $2\,\mu\mathrm{K}$.}  The logical ordering is $(|000\rangle,|001\rangle,|010\rangle,|011\rangle, |100\rangle,|101\rangle,|110\rangle,|111\rangle)$, with the blocked branch first, so that $U_{\mathrm{coh}}=B\oplus A$.  Ideally $B=I_4$, whereas $A$ is SWAP for CSWAP and iSWAP for C-iSWAP.  The notation, global-phase subtraction, eight-state target model, and decay-only definition of $P_{\mathrm{RydLos}}$ are the same as in Table~\ref{tab:two-uncontrolled-maps}.  The eight loss probabilities are ordered as the four blocked inputs followed by the four active inputs; the blocked-branch propagation includes the Rydberg excitation held on the control atom during the target pulse.}
\label{tab:two-controlled-maps}
\end{table*}

\section{Interaction-dependent light shifts and phase fluctuations}

Interaction uncertainty affects the gate by changing the light shifts
accumulated during target driving. We compare the CSWAP and
standalone SWAP operating points. The physical origin and perturbative
order of these shifts differ between the SWAP-active branch, with the
control in $|1_c\rangle$, and the blocked branch, with the control in
$|r_c\rangle$. In the latter case, excitation of a single target already
samples the control--target interaction. In the active branch, the
target--target interaction enters only through processes involving
excitation of both targets. Consequently, the interaction-dependent
light shifts generally occur at higher perturbative order in the
SWAP-active branch.

For an instantaneous dressed-state energy $E_n(t)$, the residual
phase after fixed compensation of the nominal operation is
\begin{equation}
\delta\phi_n
=
-\sum_a\delta V_a
\int_0^T
\frac{\partial E_n(t)}{\partial V_a}\,dt,
\qquad
a\in\{c1,c2,tt\}.
\label{eq:phase_noise_response}
\end{equation}
Only interaction-dependent terms contribute to this expression.
Interaction-independent light shifts can produce substantial nominal
phases but do not contribute to the interaction-noise sensitivity.

\subsection{Blocked branch: control in $|r_c\rangle$}

For $|r_c00\rangle$, the leading light shift $\Omega_1^2/2\Delta$ is independent of the interactions. The leading interaction-dependent correction is fourth order:
\begin{equation}
\Delta E_{r_c00}^{(4)}=
-\frac{\Omega_1(t)^2\Omega_2^2}{16\Delta^2}
\left(
\frac{1}{V_{c1}}+\frac{1}{V_{c2}}
\right).
\end{equation}

For $|r_c01\rangle$ and $|r_c10\rangle$, the second-order microwave
shifts through the neighboring computational states cancel.
The optical coupling produces the interaction-dependent shifts
\begin{equation}
\Delta E_{r_c01}^{(2)}
=
\frac{\Omega_2^2}{4(\Delta-V_{c2})},
\qquad
\Delta E_{r_c10}^{(2)}
=
\frac{\Omega_2^2}{4(\Delta-V_{c1})}.
\end{equation}
For $|r_c11\rangle$, other than interaction independent light shift of $\Omega_1^2/2\Delta$, the interaction dependent term has the same order 
\begin{equation}
\Delta E_{r_c11}^{(2)}
=
\frac{\Omega_2^2}{4}
\left[
\frac{1}{\Delta-V_{c1}}
+
\frac{1}{\Delta-V_{c2}}
\right].
\end{equation}
These second-order terms are the dominant source of
control--target interaction-induced phase fluctuations in the
parameter regime considered here.

\subsection{SWAP-active branch: control in $|1_c\rangle$}

The target dynamics in this branch are those of a two-qubit SWAP.
The leading optical shifts associated with a single target excitation
are independent of $V_{tt}$. Interaction sensitivity instead arises
from higher-order contributions to the dressed-state light shifts.

For $|1_c11\rangle$, the leading interaction-dependent contribution
is fourth order:
\begin{equation}
 E_{1_c11,\mathrm{int}}^{(4)}
=
\frac{\Omega_2^4}{4\Delta^2}
\frac{1}{2\Delta-V_{tt}}.
\label{eq:active_11_light_shift}
\end{equation}
Thus,
\begin{equation}
\delta\phi_{1_c11}
\simeq
-\frac{T\Omega_2^4}
{4\Delta^2(2\Delta-V_{tt})^2}\,\delta V_{tt}.
\end{equation}
The interaction-dependent shift is therefore two orders higher
than the corresponding second-order shift of $|r_c11\rangle$.

For $|1_c01\rangle$ and $|1_c10\rangle$, we use the exchange
and phase evolution derived from the effective Hamiltonian in the main text. These
expressions describe the four-photon pathway through $|11\rangle$
and $|\overline{1r}\rangle$, including the light shift of
$|\overline{1r}\rangle$ caused by virtual coupling to $|rr\rangle$.
In the dispersive limit with small Rydberg admixture, the same
pathway approximation gives the effective Hamiltonian in the basis $\{|01\rangle,|10\rangle\}$
\begin{equation}
H(t)=
\begin{pmatrix}
C+J(t)&J(t)\\
J(t)&C+J(t)
\end{pmatrix},\qquad
J(t)=\frac{\lambda_g(t)-C}{2},
\label{eq:mixed_interaction_hamiltonian}
\end{equation}
 with $A$, $B$, $C$,
and $\lambda_g$ defined in Eqs.~(4) and (7) of the main text. Consequently,
$\theta(T)=\int_0^T J(t)\,dt$ and
$\bar\phi(T)=\int_0^T[C+J(t)]\,dt$. The signed exchange
coupling $J$ is half the population Rabi frequency in magnitude
$\Omega_{\mathrm{eff}}=2|J|$.

For the active branch, $\delta_r=0$. In the regime
$|B-C|\gg\sqrt{2}|A|$, substitution of the main-text coefficients
gives
\begin{equation}
\begin{aligned}
J(t)&\simeq-\frac{A(t)^2}{B-C}
=\frac{\Omega_1(t)^2}{2\Delta}
\frac{V_{tt}-\Delta}{3V_{tt}-\Delta},
\end{aligned}
\label{eq:mixed_path_exchange_rate}
\end{equation}
Holding the drives, detuning, and duration fixed
therefore gives $\delta\bar\phi=\delta\theta$. With the
amplitude-phase convention used here,
$\phi_{\mathrm{com}}=-\bar\phi$, a small quasistatic
interaction variation produces
\begin{equation}
\begin{aligned}
\delta\theta=-\delta\phi_{\mathrm{com}}
&\simeq\frac{\delta V_{tt}}{(3V_{tt}-\Delta)^2}
\int_0^T\Omega_1(t)^2\,dt
=\frac{\kappa T\Omega_{1,\max}^2\,\delta V_{tt}}
{(3V_{tt}-\Delta)^2}.
\end{aligned}
\label{eq:mixed_path_phase_response}
\end{equation}
We have written $\Omega_1(t)=\Omega_{1,\max}f(t/T)$ and
$\kappa=\int_0^1 f(u)^2\,du$. The cancellation of $\Omega_2$ in this leading
expression does not remove the conditions for elimination, i.e. 
$|\Delta|\gg|\Omega_{1,2}|$, $|V_{tt}-\Delta|\gg|\Omega_2|$,
and $|B-C|\gg\sqrt{2}|A|$ must hold. In particular, the
apparent pole at $3V_{tt}=\Delta$ lies outside this regime.

For $|1_c00\rangle$, the leading dark-state approximation suppresses
the light shift. The shortest closed paths that sample $V_{tt}$
contain eight couplings. 

Table~\ref{tab:interaction_light_shift_comparison} summarizes the
comparison. The listed path orders describe the bare excitation
structure; they are not a uniform nondegenerate expansion of the
nearly resonant SWAP dynamics. The numerical size of each phase
response also depends on the dressed-state energy denominators.

\begin{table*}[t]
\centering
\begin{tabular}{lcc}
\hline\hline
Target input
& $|1_c\rangle$: SWAP-active branch
& $|r_c\rangle$: blocked branch\\
\hline
$|00\rangle$
& $\Delta E_{00,\mathrm{int}}$; eighth-order paths
& $-\dfrac{\Omega_1^2\Omega_2^2}{16\Delta^2}
   \left(\dfrac{1}{V_{c1}}+\dfrac{1}{V_{c2}}\right)$;
   fourth order\\[3mm]
$|01\rangle,\ |10\rangle$
& $\Delta E_{\mathrm{com,int}}=L_{tt}
   =\dfrac{1}{2}\Delta\lambda_{s,\mathrm{int}}$;
   sixth-order paths
& $\dfrac{\Omega_2^2}{4(\Delta-V_{c2})},\
   \dfrac{\Omega_2^2}{4(\Delta-V_{c1})}$;
   second order\\[3mm]
$|11\rangle$
& $\dfrac{\Omega_2^4}{4\Delta^2}
   \left[\dfrac{1}{2\Delta-V_{tt}}-\dfrac{1}{2\Delta}\right]$;
   fourth order
& $\dfrac{\Omega_2^2}{4}
   \left[\dfrac{1}{\Delta-V_{c1}}
   +\dfrac{1}{\Delta-V_{c2}}\right]$;
   second order\\
\hline\hline
\end{tabular}
\caption{Interaction-dependent light shifts for the two control
branches. Interaction-independent terms are omitted.
$\Delta\lambda_{s,\mathrm{int}}$ denotes the
interaction-dependent part of the symmetric dressed-branch energy.
The active-branch shifts of $|00\rangle$, $|10\rangle$ and $|01\rangle$ are
evaluated using the dressed-state Hamiltonian to retain collective
near-degeneracies.}
\label{tab:interaction_light_shift_comparison}
\end{table*}

\subsection{Estimated phase fluctuations and gate infidelity}

We use the rounded $150\,\mu\mathrm{K}$ operating points in
rows 2 (SWAP) and 8 (CSWAP) of Table~I of the main text.
In units of MHz for frequencies and
$\mu$s for time, the parameters are
\begin{equation}
\begin{aligned}
(\Omega_{1,\max},\Omega_2,\Delta,V_{tt})/(2\pi)
&=(85.8,234.78,1510,1017),&&\mathrm{SWAP},\\
(\Omega_{1,\max},\Omega_2,\Delta,V_{tt},V_{ct})/(2\pi)
&=(134.4,205.75,1750,1230,-2840),&&\mathrm{CSWAP},
\end{aligned}
\end{equation}
with $(T,\sigma/T)=(0.697,0.31)$ and $(0.4,0.38)$,
respectively. 
 The  Gaussian $\Omega_1$ pulse of the form
\begin{equation}
\Omega_1(t)=\Omega_{1,\max}f(t/T),\qquad
f(u)=\frac{e^{-(u-1/2)^2/(2s^2)}-e^{-1/(8s^2)}}
{1-e^{-1/(8s^2)}},\quad s=\sigma/T.
\label{eq:phase_estimate_pulse}
\end{equation}
is considered. These operating points were jointly optimized for nominal
performance and robustness, rather than chosen to minimize
only the interaction-phase response.

In the active branch, where the control atom remains in its
ground state, fluctuations of the target--target interaction
modify both the accumulated phases and the population exchange
between $|01\rangle$ and $|10\rangle$.
To evaluate these effects, we diagonalize the full target
Hamiltonian given in the main text.

We denote the relevant normalized instantaneous eigenstates by
$|\psi_j(t)\rangle$, with eigenvalues $E_j(t)$ and branch labels
$j=00,11,s,a$. We track the branches that connect at the start
of the pulse to $|00\rangle$, $|11\rangle$, and the symmetric
and antisymmetric states
\begin{equation}
|s\rangle=\frac{|01\rangle+|10\rangle}{\sqrt{2}},
\qquad
|a\rangle=\frac{|01\rangle-|10\rangle}{\sqrt{2}}.
\end{equation}
The computational input $|01\rangle$ is the normalized sum
of these two states, and $|10\rangle$ is their normalized
difference:
\begin{equation}
|01\rangle=\frac{|s\rangle+|a\rangle}{\sqrt{2}},
\qquad
|10\rangle=\frac{|s\rangle-|a\rangle}{\sqrt{2}}.
\end{equation}
Thus either computational input evolves through both branches. During the pulse, the eigenvalue difference between the symmetric and antisymmetric dressed branches generates a relative dynamical phase. When this phase reaches $\pi$ and the dressed branches return to the computational subspace, the two contributions recombine to exchange $|01\rangle$ and $|10\rangle$.

In the adiabatic approximation, the dynamical phase of each
dressed state is
$\phi_j=-\int_0^T E_j(t)\,dt$.
Since the target--target interaction enters the Hamiltonian
as $V_{tt}|rr\rangle\langle rr|$, we would have
\begin{equation}
\frac{\partial E_j(t)}{\partial V_{tt}}
=
|\langle rr|\psi_j(t)\rangle|^2.
\end{equation}
Thus, a small interaction fluctuation that is constant during
a gate produces the first-order phase change
\begin{equation}
\delta\phi_j=-A_j\zeta_{tt},
\qquad
A_j=V_{tt}\int_0^T
|\langle rr|\psi_j(t)\rangle|^2\,dt,
\qquad
\zeta_{tt}=\frac{\delta V_{tt}}{V_{tt}}.
\end{equation}
The sensitivity is therefore determined by the time-integrated
population of $|rr\rangle$ in each dressed state.

For the exchange-symmetric Hamiltonian considered here,
the antisymmetric branch has no $|rr\rangle$ component,
so $A_a=0$. The symmetric branch acquires the phase
$\delta\phi_s=-A_s\zeta_{tt}$. In the computational basis
$\{|01\rangle,|10\rangle\}$, the phase-error generator is
$\delta\Phi_{\mathrm{ex}}=\delta\phi_s|s\rangle\langle s|$:
\begin{equation}
\delta\Phi_{\mathrm{ex}}
=
-\frac{A_s\zeta_{tt}}{2}
\begin{pmatrix}
1&1\\
1&1
\end{pmatrix},
\qquad
U_{\mathrm{err}}=e^{i\delta\Phi_{\mathrm{ex}}}.
\end{equation}
The diagonal part shifts the output phases of both mixed
inputs equally, while the off-diagonal part changes the
exchange angle. Both contributions have magnitude
$|A_s\zeta_{tt}|/2$.
The common phase is common only to this two-state subspace
and can therefore contribute to phase errors relative to
$|00\rangle$ and $|11\rangle$.

\begin{table*}[t]
\centering
\begin{tabular}{lcc}
\hline\hline
Target input
& RMS phase for $|1_c\rangle$ (rad)
& RMS phase for $|r_c\rangle$ (rad)\\
\hline
$|00\rangle$
& $8.90\times10^{-5}\ /\ 4.45\times10^{-5}$
& $1.03\times10^{-4}\ /\ 6.49\times10^{-5}$\\
$|01\rangle,\ |10\rangle$
& $0.0144\ /\ 0.00718$
& $0.0359\ /\ 0.0359$\\
$|11\rangle$
& $8.56\times10^{-4}\ /\ 4.28\times10^{-4}$
& $0.0567\ /\ 0.0359$\\
\hline\hline
\end{tabular}
\caption{Estimated RMS phase fluctuations for the rounded
$150\,\mu$K CSWAP point (row 8 of main-text Table~I).
Each entry lists triangle/line. Active-branch entries follow
the instantaneous dressed energies; blocked-branch entries
retain the leading perturbative shifts. For active mixed
inputs, the listed common-phase RMS also equals the
exchange-angle RMS. These are adiabatic phase diagnostics,
not phase variances from full propagation of bare inputs.}
\label{tab:phase_rms_comparison}
\end{table*}

In the blocked branch, the control atom occupies its
Rydberg state and suppresses the target exchange.
Fluctuations of the control--target interactions nevertheless
change the light shifts and hence the phases accumulated
by the target states.
We write
\begin{equation}
V_{c1}=V_{ct}+\delta V_{c1},
\qquad
V_{c2}=V_{ct}+\delta V_{c2},
\end{equation}
where $V_{ct}$ is the nominal interaction, assumed equal
for the two targets. To leading perturbative order,
the interaction-dependent light shifts are
\begin{align}
E_{00}^{(\mathrm{int})}(t)
&\simeq
-\frac{\Omega_1(t)^2\Omega_2^2}{16\Delta^2}
\left(\frac{1}{V_{c1}}+\frac{1}{V_{c2}}\right),\\
E_{01}^{(\mathrm{int})}
&\simeq \frac{\Omega_2^2}{4(\Delta-V_{c2})},\\
E_{10}^{(\mathrm{int})}
&\simeq \frac{\Omega_2^2}{4(\Delta-V_{c1})},\\
E_{11}^{(\mathrm{int})}
&\simeq \frac{\Omega_2^2}{4}
\left(\frac{1}{\Delta-V_{c1}}
+\frac{1}{\Delta-V_{c2}}\right).
\end{align}
Terms independent of these interactions are omitted because
they do not contribute to the responses considered here.

Expanding these shifts to first order about
$V_{c1}=V_{c2}=V_{ct}$ and using
$\delta\phi=-\int_0^T\delta E(t)\,dt$ gives
\begin{align}
\delta\phi_{00}^{(\mathrm{blocked})}
&\simeq
-\frac{\Omega_2^2}{16\Delta^2V_{ct}^2}
\left[\int_0^T\Omega_1(t)^2\,dt\right]
(\delta V_{c1}+\delta V_{c2}),\\
\delta\phi_{01}^{(\mathrm{blocked})}
&\simeq
-\frac{T\Omega_2^2}{4(\Delta-V_{ct})^2}
\,\delta V_{c2},\\
\delta\phi_{10}^{(\mathrm{blocked})}
&\simeq
-\frac{T\Omega_2^2}{4(\Delta-V_{ct})^2}
\,\delta V_{c1},\\
\delta\phi_{11}^{(\mathrm{blocked})}
&\simeq
-\frac{T\Omega_2^2}{4(\Delta-V_{ct})^2}
(\delta V_{c1}+\delta V_{c2}).
\end{align}
Here the subscripts specify the target states, with the
control in its Rydberg state throughout this part of the
sequence. These expressions assume a constant $\Omega_2$
and interaction fluctuations that are constant during
the gate. They give the additional phases relative to
the nominal operation, with the nominal phase corrections
held fixed. Their signs must be preserved when evaluating
relative-phase errors between computational states,
including those belonging to different control branches.

These estimates isolate the light-shift mechanism. The
finite pulses also populate other dressed components, so
their full noisy dynamics need not be described by a single
adiabatic branch. In particular, the small standalone
phase-model values cannot be used to infer a negligible
interaction error in the complete SWAP operation. Full
propagation, as in Fig.~4 of the main text, is needed to
assess the total sensitivity.

\section{Intensity-noise sensitivity and coupling ratios}
\label{sec:intensity_ratios}

Intensity fluctuations perturb both the exchange angle and the
relative phases of the computational components. The $\Omega_1$
drive controls the exchange area, whereas the leading optical
light shifts already depend on $\Omega_2$ at second order.
Their accumulated effect depends on the actual gate duration,
as well as the coupling ratios, detunings, and interactions.
The numerical estimates below use the same rounded $150\,\mu$K
Table~I parameters as the preceding section.

\subsection{Leading optical phase response}

For a small fractional intensity fluctuation
$\epsilon_2=\delta I_2/I_2$ of one excitation field, with the
other field and the effective detuning fixed,
$\delta\Omega_2/\Omega_2\simeq\epsilon_2/2$.
For constant $\Omega_2$ during target driving, define
\begin{equation}
a=-\frac{T\Omega_2^2}{4\Delta},\qquad
b=-\frac{T\Omega_2^2}{4(\Delta-V_{ct})}.
\label{eq:optical_phase_coefficients}
\end{equation}
With the convention $\phi=-\int E\,dt$, the leading optical
phases in the target basis $(00,01,10,11)$ are
\begin{equation}
\boldsymbol\phi_{1_c}^{(2)}=(0,a,a,2a),\qquad
\boldsymbol\phi_{r_c}^{(2)}=(0,b,b,2b).
\label{eq:leading_optical_phases}
\end{equation}
The active-branch expression also applies to standalone SWAP.
The mixed-input entries are common phases; additional
exchange-dependent contributions are omitted at this order.
The zero entry for $|00\rangle$ uses the leading dark-state
approximation. Microwave-induced phases and higher-order
optical corrections are not included in these expressions.

For common quasistatic intensity noise on both targets,
$\delta\boldsymbol\phi\simeq\epsilon_2\boldsymbol\phi^{(2)}$.
Equivalently, writing $\eta_2=\delta\Omega_2/\Omega_2$ gives
$\delta\boldsymbol\phi\simeq2\eta_2\boldsymbol\phi^{(2)}$.
At fixed duration, for example,
\begin{equation}
\frac{\partial\phi_{1_c01}^{(2)}}{\partial\Omega_2}
=-\frac{T\Omega_2}{2\Delta},\qquad
\frac{\partial\phi_{r_c01}^{(2)}}{\partial\Omega_2}
=-\frac{T\Omega_2}{2(\Delta-V_{ct})},
\end{equation}
with twice these derivatives for the corresponding $|11\rangle$
inputs. Fixed nominal phase corrections do not remove these
random relative phases.

For Gaussian fractional intensity noise of RMS amplitude
$s_I=\sigma(\delta I_2/I_2)$, the leading small-noise estimates are
\begin{equation}
\begin{aligned}
1-\overline F_{\mathrm{phase}}^{\mathrm{SWAP}}
&\simeq\frac{2}{5}a^2s_I^2,\\
1-\overline F_{\mathrm{phase}}^{\mathrm{CSWAP}}
&\simeq\frac{4}{9}(a^2+b^2-ab)s_I^2.
\end{aligned}
\label{eq:intensity_phase_infidelity}
\end{equation}
The second expression assumes that the same intensity
fluctuation acts in both control branches. These estimates
assume ideal population dynamics and retain only the leading
optical phase response. They do not include exchange-angle
errors, nonadiabatic transitions, leakage, or decay.

\begin{table*}[t]
\centering
\begin{tabular}{lcc}
\hline\hline
Quantity & SWAP (Table~I, row 2) & CSWAP (Table~I, row 8)\\
\hline
$T$ ($\mu$s) & $0.697$ & $0.400$\\
$\Omega_2/\Omega_{1,\max}$ & $2.7364$ & $1.5309$\\
$\Omega_2/\Delta$ & $0.15548$ & $0.11757$\\
$a$ (rad) & $-39.9667$ & $-15.1992$\\
$b$ (rad) & --- & $-5.79491$\\
$1-\overline F_{\mathrm{phase}}$ at $s_I=0.2\%$
& $0.2556\%$ & $0.03138\%$\\
\hline\hline
\end{tabular}
\caption{Leading intensity-phase estimates using the actual
rounded durations and couplings in rows 2 and 8 of Table~I of
the main text. The infidelities use the small-noise expressions
in Eq.~\eqref{eq:intensity_phase_infidelity}, rather than a full
noisy propagation.}
\label{tab:intensity_current_parameters}
\end{table*}

\section{Operation regime for joint intensity and interaction robustness}
\label{sec:joint_regime}

The preceding discussion identifies how the original driving and
interaction parameters can be chosen to reduce phase sensitivity
while preserving the exchange. We take $\Delta>0$, as in Table~I,
and use the pulse shape already defined above,
$\Omega_1(t)=\Omega_{1,\max}f(t/T)$ with
$\kappa=\int_0^1f(u)^2\,du$.
The exchange coupling in Eq.~\eqref{eq:mixed_path_exchange_rate}
gives
\begin{equation}
\theta(T)\simeq
\frac{\kappa T\Omega_{1,\max}^2}{2\Delta}
\frac{V_{tt}-\Delta}{3V_{tt}-\Delta}.
\label{eq:regime_exchange_area}
\end{equation}
For the first full SWAP, $|\theta(T)|=\pi/2$, the corresponding
dispersive estimate of the duration is
\begin{equation}
T\simeq\frac{\pi\Delta}{\kappa\Omega_{1,\max}^2}
\left|\frac{3V_{tt}-\Delta}{V_{tt}-\Delta}\right|.
\label{eq:dispersive_duration}
\end{equation}
This estimate includes the $|rr\rangle$ light shift and uses
$\Omega_{\mathrm{eff}}=2|J|$. It is used below to compare
calibrated pulse families; all numerical estimates retain the
actual optimized durations in Table~I. During a noise realization,
the duration, detuning, pulse shape, and nominal phase corrections
remain fixed.

\subsubsection{Intensity sensitivity and coupling ratios}

For a fractional intensity fluctuation of one excitation field,
with the other excitation parameters fixed,
$\delta\Omega_j/\Omega_j=\delta I_j/(2I_j)$.
The leading optical phase responses are
\begin{equation}
\begin{aligned}
\delta\phi_{1_c01}^{(2)}\big|_{I_2}
&\simeq-\frac{T\Omega_2^2}{4\Delta}\frac{\delta I_2}{I_2},\\
\delta\phi_{r_c01}^{(2)}\big|_{I_2}
&\simeq-\frac{T\Omega_2^2}{4(\Delta-V_{ct})}
\frac{\delta I_2}{I_2}.
\end{aligned}
\label{eq:regime_intensity_responses}
\end{equation}
The same expressions apply to $|10\rangle$, and the leading
$|11\rangle$ optical responses are twice as large for equal
control--target interactions. Substituting the calibrated duration
into the active response gives
\begin{equation}
\delta\phi_{1_c01}^{(2)}\big|_{I_2}
\simeq-\frac{\pi}{4\kappa}
\left(\frac{\Omega_2}{\Omega_{1,\max}}\right)^2
\left|\frac{3V_{tt}-\Delta}{V_{tt}-\Delta}\right|
\frac{\delta I_2}{I_2}.
\end{equation}
Thus, at fixed pulse shape and $V_{tt}/\Delta$, reducing
$\Omega_2/\Omega_{1,\max}$ decreases the leading phase
sensitivity quadratically. Increasing $\Delta$ alone can require a longer
pulse, leaving $T\Omega_2^2/\Delta$ in Eq.~\eqref{eq:regime_intensity_responses} unchanged.
For the blocked branch, an attractive interaction $V_{ct}<0$
reduces the optical phase response by the additional factor
$\Delta/(\Delta+|V_{ct}|)$.

The ratio $\Omega_2/\Delta$ controls higher-order corrections.
For the pair-mediated fourth-order phase of $|1_c11\rangle$,
\begin{equation}
\delta\phi_{1_c11,\mathrm{int}}^{(4)}\big|_{I_2}
\simeq-\frac{T\Omega_2^4}{2\Delta^2(2\Delta-V_{tt})}
\frac{\delta I_2}{I_2}.
\end{equation}
Its intensity derivative relative to that of the leading
$|1_c11\rangle$ optical phase is
$\Omega_2^2/[\Delta(2\Delta-V_{tt})]$, equal to $0.0182$
and $0.0107$ for the current SWAP and CSWAP points.
Reducing $\Omega_2/\Delta$ suppresses this correction provided
$V_{tt}$ remains away from $2\Delta$.
Interaction-independent higher-order shifts also contribute
to the complete intensity response.

The leading exchange coupling is independent of $\Omega_2$
within this dispersive approximation, but it remains sensitive
to the drive $\Omega_1$:
\begin{equation}
\delta\theta\big|_{I_1}\simeq\theta(T)\frac{\delta I_1}{I_1}.
\end{equation}

\subsubsection{Interaction sensitivity and distance from resonances}

At fixed drives and duration, the leading interaction responses are
\begin{equation}
\begin{aligned}
\delta\phi_{r_c01}\big|_{V_{ct}}
&\simeq-\frac{T\Omega_2^2}{4(\Delta-V_{ct})^2}\,\delta V_{ct},\\
\delta\phi_{1_c11}^{(4)}\big|_{V_{tt}}
&\simeq-\frac{T\Omega_2^4}{4\Delta^2(2\Delta-V_{tt})^2}
\,\delta V_{tt},\\
\delta\theta\big|_{V_{tt}}
=-\delta\phi_{\mathrm{com}}\big|_{V_{tt}}
&\simeq\frac{\kappa T\Omega_{1,\max}^2}
{(3V_{tt}-\Delta)^2}\,\delta V_{tt}.
\end{aligned}
\label{eq:regime_interaction_responses}
\end{equation}
The blocked $|11\rangle$ phase response is the sum of the two
control--target contributions, whose correlations must be retained.
For a nonzero calibrated exchange angle,
\begin{equation}
\frac{\delta\theta}{\theta(T)}\bigg|_{V_{tt}}
\simeq\frac{2\Delta\,\delta V_{tt}}
{(V_{tt}-\Delta)(3V_{tt}-\Delta)}.
\label{eq:fractional_exchange_sensitivity}
\end{equation}
These expressions identify the interaction regions to avoid.
Near $3V_{tt}=\Delta$, the active dressed-state gap becomes small.
Approaching $V_{tt}=\Delta$ encounters an eliminated-state resonance
and suppresses the approximate exchange rate, increasing the duration
required for a fixed gate angle. The $|11\rangle$ phase response
is enhanced near $V_{tt}=2\Delta$, while the blocked optical
response is enhanced near $V_{ct}=\Delta$.

At fixed duration, reducing $\Omega_2$ suppresses the leading
control--target phase response as $\Omega_2^2$ and the displayed
active $|11\rangle$ response as $\Omega_2^4$. However, the
leading $V_{tt}$ exchange-angle response is independent of
$\Omega_2$ and must also be controlled through $V_{tt}/\Delta$
and the calibrated duration.
A sufficient asymptotic regime is $|V_{tt}|\gg\Delta$ and,
for controlled gates, $|V_{ct}|\gg\Delta$, while avoiding all
intermediate- and pair-state resonances. For example,
\begin{equation}
\left|\frac{\delta\theta}{\theta(T)}\right|
\simeq\frac{2\Delta}{3|V_{tt}|}
\left|\frac{\delta V_{tt}}{V_{tt}}\right|,\qquad
|\delta\phi_{r_c01}|
\simeq\frac{T\Omega_2^2}{4|V_{ct}|}
\left|\frac{\delta V_{ct}}{V_{ct}}\right|.
\end{equation}
The displayed interaction responses therefore decrease as $1/|V|$
at fixed fractional uncertainty, or as $1/V^2$ at fixed absolute
uncertainty. Strong control--target interaction also suppresses
the blocked intensity response, but does not suppress the active
optical phase. The present optimized points balance these
sensitivities at finite interaction strengths; the asymptotic
regime is not a requirement for their operation.

\subsubsection{Preserving exchange while reducing sensitivity}

The ratio $\Omega_2/\Omega_{1,\max}$ cannot be reduced
arbitrarily. Using the previously defined $A$, $B$, and $C$,
weak Rydberg admixture in the active branch requires
\begin{equation}
\frac{\sqrt{2}|A(t)|}{|B-C|}
=2\left|\frac{\Omega_1(t)}{\Omega_2}\right|
\left|\frac{V_{tt}-\Delta}{3V_{tt}-\Delta}\right|\ll1.
\end{equation}
This provides a lower limit on $\Omega_2/\Omega_{1,\max}$,
whereas the intensity-phase response favors a smaller ratio.
For a specified RMS intensity uncertainty, the leading active
optical phase remains small when
\begin{equation}
\frac{T\Omega_2^2}{4\Delta}
\sigma\!\left(\frac{\delta I_2}{I_2}\right)\ll1.
\end{equation}
The interaction-induced phases and exchange-angle variations in
Eq.~\eqref{eq:regime_interaction_responses} must likewise remain
small over the actual gate duration.

The dispersive conditions on the original parameters are
\begin{equation}
\begin{aligned}
\Delta&\gg|\Omega_{1,2}|,\qquad
|V_{tt}-\Delta|\gg|\Omega_2|,\\
|2\Delta-V_{tt}|&\gg\frac{\Omega_2^2}{\Delta}.
\end{aligned}
\end{equation}
For the blocked branch, the corresponding conditions include
\begin{equation}
\begin{aligned}
|\Delta-V_{ct}|&\gg|\Omega_2|,\qquad
|V_{tt}+V_{ct}-\Delta|\gg|\Omega_2|,\\
|2\Delta-V_{tt}-2V_{ct}|&\gg
\frac{\Omega_2^2}{|\Delta-V_{ct}|}.
\end{aligned}
\end{equation}
The last inequalities in these two displays compare pair-state
detunings with the induced two-photon couplings and light shifts.
The blocked branch must also have $|B-C|\gg\sqrt{2}|A|$ and
accumulate negligible exchange. When $B-C\simeq V_{ct}$,
\begin{equation}
\left|\frac{J_{r_c}}{J_{1_c}}\right|
\simeq\frac{\Omega_2^2}{4|V_{ct}\Delta|}
\left|\frac{3V_{tt}-\Delta}{V_{tt}-\Delta}\right|\ll1.
\end{equation}

For constant $B$ and $C$, adiabatic following of the active dressed
branch additionally requires
\begin{equation}
\frac{\sqrt{2}|(B-C)\dot A(t)|}
{[(B-C)^2+8A(t)^2]^{3/2}}\ll1.
\end{equation}
Reducing $\Omega_2$ shrinks the dressed gap and can violate this
condition. Lengthening the pulse to restore adiabaticity increases
Rydberg exposure. The desired regime therefore combines a reduced
$\Omega_2/\Omega_{1,\max}$ that still preserves collective exchange,
a small $\Omega_2/\Delta$, interactions away from the specified
resonances, and a duration that balances phase errors and decay.
The parameters listed in Table I of the main text are rounded values obtained by numerical optimization to minimize sensitivity to Doppler shifts and fluctuations in laser intensity and interatomic interactions, while accounting for Rydberg decay and population-transfer and phase errors.

\section{Temporal correlations and numerical scope}

Only residual relative phases affect the gate fidelity.
For time-dependent common intensity noise, their linear
response is
\begin{equation}
\delta\phi_i-\delta\phi_j
=-\int_0^T[g_i(t)-g_j(t)]\epsilon_2(t)\,dt,
\qquad
g_i(t)=\frac{\Omega_2(t)}{2}
\frac{\partial E_i(t)}{\partial\Omega_2}.
\end{equation}
For covariance
$C_\epsilon(t,t')=\langle\epsilon_2(t)\epsilon_2(t')\rangle$,
\begin{equation}
\begin{aligned}
\operatorname{Var}(\delta\phi_i-\delta\phi_j)
={}&\int_0^Tdt\int_0^Tdt'\,[g_i(t)-g_j(t)]\\
&\times[g_i(t')-g_j(t')]C_\epsilon(t,t').
\end{aligned}
\end{equation}
Here $i,j$ label the modes of the effective computational
evolution, including the exchange block. The technical-noise sweeps
in Fig.~4 of the main text use quasistatic Gaussian
fluctuations, held fixed during each gate realization.
A prediction for faster noise requires its temporal
covariance or spectrum. For two-photon excitation, any
associated differential AC Stark-shift fluctuations must
also be included if they change the effective detuning.


\begin{thebibliography}{100}
\bibitem{explanation2} The coupling $\Omega_1$ may be implemented by a two-photon Raman transition between the hyperfine qubit states. For sufficiently large intermediate-state detuning, adiabatic elimination yields an effective coupling of the same form; matching its amplitude, phase, and detuning, while compensating differential AC Stark shifts, preserves the blockade-enabled SWAP mechanism.
\bibitem{Explain} The goal of Fig. 3 is to quantify the operating regime of parameters that lead to desired population rotation. Adjusting relative phases requires fine-tuning of controlling parameters and is a computationally extensive task. Hence, we use rotation fidelity obtained from Eq.~\eqref{FidDefenition} by replacing $M$ with $M_{\mathrm{rot}}=|U_{\mathrm{id}}|^\dagger|U_{\mathrm{gate}}|$, where absolute values are taken elementwise. It quantifies errors in transition-amplitude magnitudes, including leakage, while discarding relative phases.
\end{thebibliography}
\end{document}